\documentclass[pdflatex,sn-mathphys-ay]{sn-jnl}

\usepackage{amsmath,amssymb,amsfonts}%
\usepackage{amsthm}%
\usepackage{mathrsfs}%
\usepackage[title]{appendix}%
\usepackage{xcolor}%
\usepackage{textcomp}%
\usepackage{manyfoot}%
\usepackage{booktabs}%
\usepackage{listings}%
\usepackage{psfrag,epsf}
\usepackage{enumerate}
\usepackage{url} 
\usepackage{verbatim,color}
\usepackage{graphicx}
\usepackage{setspace}
\usepackage{array}
\usepackage{multirow}
\usepackage{caption}
\usepackage{enumitem}
\usepackage{algpseudocode}
\usepackage{array}
\usepackage{float}
\usepackage{caption}
\usepackage{subfigure}
\usepackage{geometry}
\usepackage{csquotes}
\usepackage{geometry}
\usepackage{bm}
\usepackage{rotating}
\usepackage{hhline}
\usepackage[linesnumbered,boxed,ruled]{algorithm2e}
\usepackage{titlecaps}
\begin{document}

\title[Article Title]{\fontsize{14}{16}\selectfont  A Multifacet Hierarchical Sentiment-Topic Model with Application to Multi-Brand Online Review Analysis}

\author[1,2]{\fnm{Qiao} \sur{Liang}}

\author*[3]{\fnm{Xinwei} \sur{Deng}}\email{xdeng@vt.edu}

\affil[1]{\orgdiv{Joint Laboratory of Data Science and Business Intelligence}, \orgname{Southwestern University of Finance and Economics}, \orgaddress{\city{Chengdu}, \country{China}}}

\affil[2]{\orgdiv{School of Statistics}, \orgname{Southwestern University of Finance and Economics}, \orgaddress{\city{Chengdu}, \country{China}}}

\affil[3]{\orgdiv{Department of Statistics}, \orgname{Virginia Tech}, \orgaddress{\city{Blacksburg}, \state{VA}, \country{United States of America}}}


\abstract{Multi-brand analysis based on review comments and ratings is a commonly used strategy to compare different brands in marketing.
It can help consumers make more informed decisions and help marketers understand their brand's position in the market.
In this work, we propose a multifacet hierarchical sentiment-topic model (MH-STM) to detect brand-associated sentiment polarities towards multiple comparative aspects from online customer reviews.
The proposed method is built on a unified generative framework that explains review words with a hierarchical brand-associated topic model and the overall polarity score with a regression model on the empirical topic distribution.
Moreover, a novel hierarchical P\'olya urn (HPU) scheme is proposed to enhance the topic-word association among topic hierarchy, such that the general topics shared by all brands are separated effectively from the unique topics specific to individual brands. 
The performance of the proposed method is evaluated on both synthetic data and two real-world review corpora.
Experimental studies demonstrate that the proposed method can be effective in detecting reasonable topic hierarchy and deriving accurate brand-associated rankings on multi-aspects.}


\maketitle

\section{Introduction}
\label{sec1}

As online activities and intelligence advance, 
an increasing amount of online reviews provide a practical way to understand customers' feedback to multiple aspects of online products and services.
Typically, a customer tends to express positive or negative opinions on different aspects of a product type or brand.
The multifacet information in reviews enables a data-driven comparative analysis of different online entities, including but not limited to different products \citep{he2023novel}, brands \citep{alzate2022mining}, platforms \citep{zhang2019examining}, and even cultures \citep{brand2022cultural}.
The primary interest of this study is to conduct a thorough modeling and analysis on the customer reviews for automatic multi-brand comparison, namely, to compare multifacet features and advantages of different brands in marketing. Note that the proposed method can be extended to other review-based multifacet comparison across different entities of interest.

\begin{figure}[ht]
\centering
\includegraphics[scale=0.45]{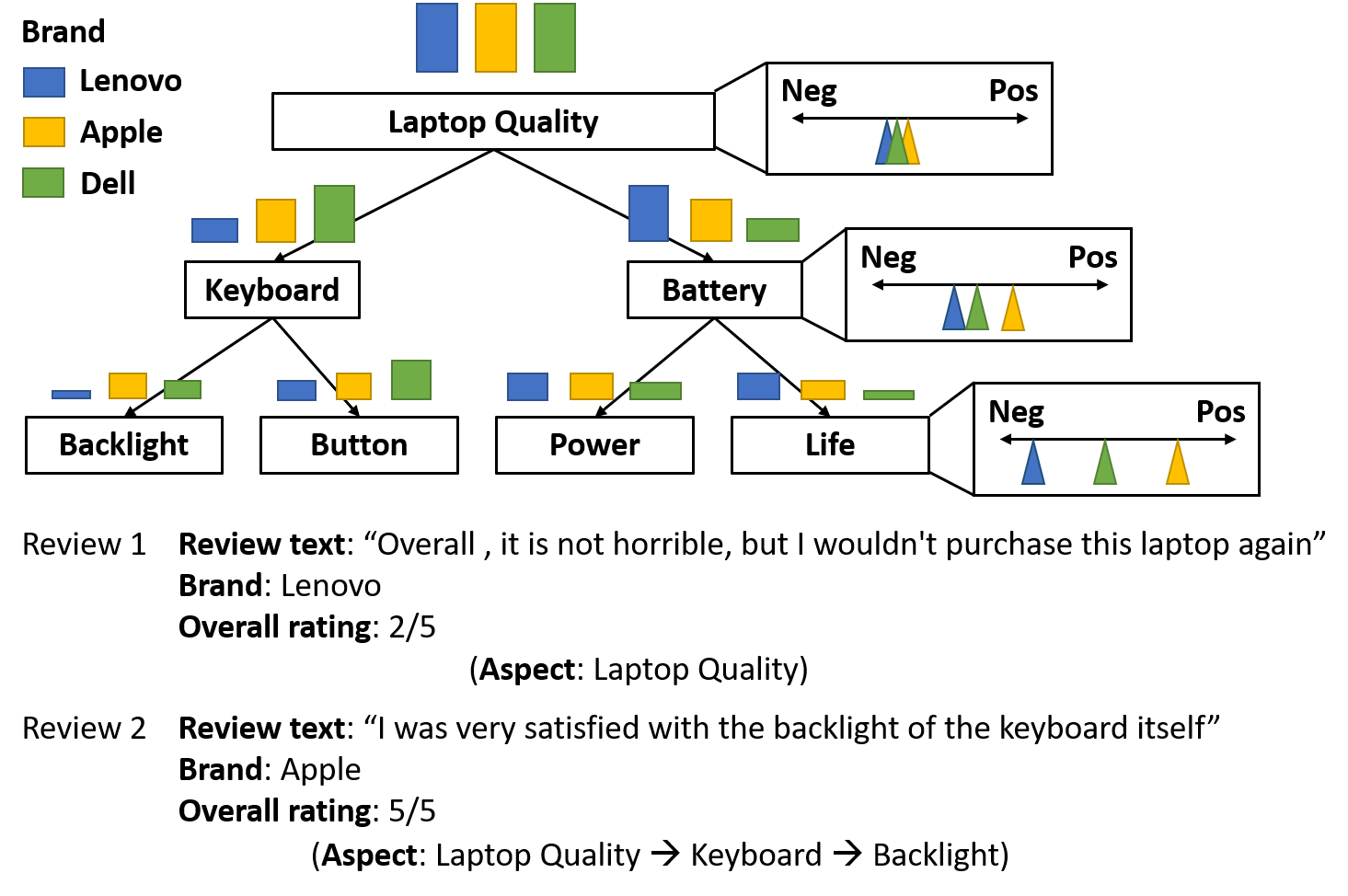}
\caption{Illustration of laptop reviews with respect to various brands. Each review is composed of a plain text and an overall rating. Different brands share multiple comparative aspects represented by a tree structure of product aspect nodes with respective distributions on that, where each aspect node is linked to multiple brand-specific sentiment polarities that influence the overall ratings in brand reviews.}
\label{fig-intro-1}
\end{figure}

Overall, the existing researches on brand aspect detection from online reviews are mainly built on classical topic models, where all topics are organized in a flat structure that assumes equal information granularities among topics. 
However, in practice, the aspects embedded in reviews tend to demonstrate different granularities, which naturally form a tree structure.
For example, Figure~\ref{fig-intro-1} illustrates online customer reviews on several laptop brands. Each review is tagged with a certain brand and composed of a plain text that describes customer's opinions on multi-aspects of the bought laptop as well as an overall rating that indicates the general sentiment polarity.
The multi-aspect information in brand-associated reviews can be captured by a tree structure of product aspect nodes shared by all brands, where different brands present different distributions on the aspect nodes, and each aspect node is linked to multiple brand-specific sentiment polarities that summarize the brand competitive performance on that aspect and influence the overall ratings observed in brand reviews.

For the purpose of multi-brand comparison, a hierarchical structure of comparative aspects rather than its flat counterpart is preferred, on account of enhanced brand differentiation with deepened levels in the hierarchy. 
A flat structure assumes the global background and local ratable aspects in reviews are mixed up, such that the local aspects are smoothed by their shared background \citep{titov2008modeling}.
In contrast, a hierarchical structure presents a varying degree of smoothness among multi-level aspects, where upper aspects of hierarchy summarize product background and general property shared among all brands, and deeper aspects capture fine-grained characteristics specific to individual brands. 
Therefore, the brand-associated strengths and weaknesses can be degeneralized from their shared background by digging into the hierarchy, which enables a salient comparison.

We propose a multifacet hierarchical sentiment-topic model (MH-STM) for automatic multi-aspect brand comparison based on the brand-associated review corpora including both review texts and overall ratings, as illustrated in Figure \ref{fig-intro-1}. 
The proposed MH-STM employs a unified framework that combines the extraction of hierarchical aspects from reviews and acquiring of brand-associated sentiment scores specific to multi-aspects. 
For deriving the hierarchical comparative aspects among brands, 
the MH-STM borrows the idea of hierarchical topic models (HTMs; \citealt{liu2016overview}) that are natural extension of flat topic models.
In the MH-STM, we adopt the hierarchical topic assumptions of hierarchical latent Dirichlet allocation (hLDA; \citealt{blei2010nested}) for the tree-structured product aspects as in Figure \ref{fig-intro-1}, and brand-associated reviews are distributed respectively over the shared topic tree.

However, a challenge of applying hLDA and its variants to our problem directly is that the detection of topic hierarchy in these models relies only on word co-occurrence rules without considering the general-to-specific characteristics of words explicitly. Some researches \citep{kang2012transfer,xu2018hierarchical} have shown that hLDA could produce unreasonable hierarchy. For example, some semantically general words are assigned with the topics of lower levels if those words do not occur frequently enough. In contrast, some semantically specific but frequently occurring words are wrongly assigned to topics of upper levels.
This mismatch brings more uncertainties in differentiating the brand-associated properties from their shared background.
To address this challenge, we develop a hierarchical P\'olya urn (HPU) scheme that can be properly applied to hierarchical topic models for enhancing the topic level assignments of words based on their general-to-specific semantics, namely, a general word is more bursty around the root topic, and a specific word is more bursty around a leaf topic. 

Overall, the major contributions of this work are two-fold. 
First, we propose a multifacet hierarchical sentiment-topic model that detects brand-associated comparative aspects in hierarchy and customers' aspect-specific polarities towards different brands jointly from online reviews. 
Compared to other multi-brand models, it is constructed on a hierarchical structure of multiple product aspects for enhanced brand differentiation with deepened levels in hierarchy.
Second, we introduce a novel hierarchical P\'olya urn (HPU) scheme in the inference of topic hierarchy among multiple brands. By incorporating different burstiness of review words in topic hierarchy according to their general-to-specific characteristics, it improves the separation of general aspects shared by all brands from the unique aspects specific to individual brands, resulting in a more accurate multi-brand comparison. 
The proposed model produces a ranked list of brands on various customer-concerned aspects that enables a straightforward brand comparison.

The outline of this article is organized as follows.
Section \ref{literature} provides a literature review in the related field.
Section \ref{method} details the proposed MH-STM with inference via a hierarchical P\'olya urn scheme. 
Section \ref{simulation} elaborates performance of the proposed method through simulation. 
Section \ref{case} evaluates model performance using two real-world online review corpora. 
We conclude this work with discussions in Section \ref{conclusion}.

\section{Literature review}
\label{literature}
\subsection{Review-based brand comparison}
Review-based comparison of multiple brands can help consumers make more informed purchasing decisions and help marketers understand their brand's position in the market \citep{colladon2018semantic,mitra2020obim}. 
Recent researches mainly aim to derive brand-associated polarities from user-generated reviews. 
Several works considered a two-stage procedure \citep{zhang2015dynamic,barry2018alcohol,sajid2022using}, which first detects brand-associated evaluation aspects from reviews by implementing probabilistic topic models \citep{lda, vayansky2020review}, and subsequently tracks customers' orientations to the extracted aspects or topics of different brands via sentiment analysis. 
In the first stage, the hierarchical topic models (HTMs; \citealt{liu2016overview}) can be applied to derive the hierarchical comparative aspects among brands.
Commonly used HTMs include two mainstream structures: Hierarchical latent Dirichlet allocation (hLDA; \citealt{blei2010nested}) and hierarchical Pachinko allocation model (hPAM; \citealt{hPAM}). 
Specifically, hLDA defines a tree-structure hierarchy where each child topic node has only one parent, and hPAM is built on a directed acyclic multilevel graph where a lower-level topic correlates with multiple upper-level topics.
An obvious limitation of the two-stage method is that the topic extraction is unsupervised and separated from the subsequent sentiment analysis, leading to reduced predictive power for customers' brand-associated polarities \citep{slda}.

Joint sentiment-topic modeling of review texts and ratings provides a powerful solution to the problem of two-stage methods \citep{Tech2023}.
There are also several works using a joint approach that derives brand-associated comparative aspects and customer polarities towards the brands simultaneously. 
For example, the text-based ideal point (TBIP) model \citep{vafa2020text} is constructed based on an unsupervised Poisson factorisation  to infer brand-level polarity scores that influence the word distributions of polarity-bearing topics shared among brands. Moreover, the brand topic model (BTM) by \citet{zhao-etal-2021-adversarial} extended the TBIP model by incorporating supervision from the document-level sentiment labels, leading to improved performance in brand ranking. And \citet{zhao2023tracking} further proposed a dynamic variant of BTM for tracking the latent brand polarity scores over time. 
However, these joint models often focus on the overall brand polarity comparison. They do not provide brand rankings under multiple aspects directly, which is less informative for the decision making of brand market strategies.

\subsection{P\'olya urn scheme in topic modeling}
The simple P\'olya urn scheme (SPU) can be naturally combined with topic modeling for enhancing the topic-word association \citep{mahmoud2008polya}.
The SPU scheme presents a self-reinforcing property known as “the rich get richer”, which helps to capture the word burstiness in a document, i.e., if a word appears once, it is more likely to appear again \citep{doyle2009accounting}.
This property enables SPU to show superior performance in the topic modeling of text documents \citep{wordburstiness}.
Recently, some enhanced P\'olya urn schemes have been proposed for improving the topic-word associations in topic inference, such as generalized P\'olya urn (GPU; \citealt{gpu-li2016topic}) and weighted P\'olya urn (WPU; \citealt{wpu-wang2020optimising}). But these methods are mainly designed for non-hierarchical topic models. 

For adaption to hierarchical topic relations, \citet{xu2018hierarchical} combined GPU with hierarchical topic sampling to produce a topic hierarchy with improved coherence and reasonable structure. 
However, it relies on the prior knowledge extracted from multiple domains' corpora.
\citet{Liang2023} developed an enhanced P\'olya urn scheme by adapting the sampling weights of words to their general-to-specific characteristics in hierarchy without requiring any external knowledge. However, it still adopts equal burstiness of words across hierarchy following the assumption of SPU.
In comparison, our proposed HPU scheme considers different burstiness of words in topic hierarchy relying only on their general-to-specific characteristics, namely, a general word is more bursty around the root topic, and a specific word is more bursty around a leaf topic.

\section{A multifacet hierarchical sentiment-topic model} 
\label{method}
For extracting comparative aspects among different brands as well as brand-associated sentiment scores for multi-aspects, Section \ref{method-1} introduces a 
multifacet hierarchical sentiment-topic model (MH-STM), and Section \ref{method-2} combines it with a novel hierarchical P\'olya urn (HPU) scheme for enhancing the topic-word association in topic hierarchy. Finally, Section \ref{method-3} demonstrates a stochastic EM algorithm for model inference.

\subsection{Model formulation}
\label{method-1}
Consider the data consisting of product reviews each indexed by $d\in\{1,2,\cdots,D_b\}$ belonging to each of competitive brands $b\in\{1,2,\cdots,B\}$ in a market. 
Without loss of generality, each review $d$ contains a normalized overall polarity score $y_d\in [0,1]$ and $S_d$ review sentences each indexed by $s\in \{1,2,\cdots,S_d\}$. Suppose a review sentence $s$ in review $d$ contains $N_{d,s}$ words denoted as $\{w_{d,s,n}\}_{n=1}^{N_{d,s}}$, thus having the total number of words in review $d$ denoted as $N_d = \sum_{s=1}^{S_d}N_{d,s}$, and each word in the observed sentence is assumed to be from the vocabulary indexed by $v\in\{1,2,\cdots,V\}$.

The MH-STM employs a unified frame of probabilistic generative process that explains the observed words and overall polarity scores in brand-associated reviews jointly. 
As illustrated in Figure \ref{fig-intro-1}, the reviews are explained by mixtures of latent topics that are organized in a tree with a pre-defined depth $L$ and unbounded branches. Each topic node $k$ is represented by a $V$-dimension multinomial word distribution $\boldsymbol{\phi}_k$ drawn from a symmetric Dirichlet prior: $\boldsymbol{\phi}_k\sim \text{Dir}(\boldsymbol{\eta})$, where $\boldsymbol{\eta}=(\eta,\cdots,\eta)^T$.  
Overall, a topic located deeper in the tree represents a more fine-grained aspect that customers are concerned with.

Given a topic node $k$ in the tree, all direct child nodes of $k$ compose its child set $\Lambda(k)$ that corresponds to each subdivision of the topic, e.g., $\Lambda(Keyboard)=\{Backlight, Button\}$ in Figure \ref{fig-intro-1}.
And $k$'s ancestor set ${\rm V}(k)$ is composed of all other nodes on the way from the node $k$ to the root, e.g., ${\rm V}(Backlight)=\{Keyboard, Laptop Quality\}$ in Figure \ref{fig-intro-1}. 
A path is defined as a sequence of distinct topic nodes connected by parent-child edges in the tree: $\boldsymbol{c}=(c_1,c_2,\cdots, c_L)^T, c_l \in \Lambda(c_{l-1}), \forall l \in \{2,\cdots,L\}$, where $c_1$ is the root node, and $c_L$ is a leaf node. Each path denotes a coherent semantic theme of reviews with different information granularities captured by each level's node in the path, e.g., $\boldsymbol{c} = (LaptopQuality,Keyboard,Button)^T$ in Figure \ref{fig-intro-1}.

To organize latent topics in an unbounded tree, we employ the nested Chinese restaurant process (nCRP; \citealt{hlda}) that is widely used in hierarchical topic modeling to place a nonparametric prior partitions on possible topic trees. 
It is assumed that different brands share the same hierarchical structure of topic nodes but with respective proportions over the nodes, which means each brand's reviews are partitioned by their respective nCRPs.
The partitioning process of nCRP specific to each brand is achieved by recursively traversing the entire tree from a specified root node.
For example, let $c_1$ be the root topic and selected as the beginning node, once given the previous $i-1$ document assignments of brand $b$, the $i$th subsequent document of brand $b$ traverses each level of tree by nCRP with parameter $\gamma$, resulting a path of topics $\boldsymbol{c}=(c_1,\cdots,c_{L})^T \sim \text{nCRP}(\gamma,b)$. 
Specifically, for each level $l\in \{2,\cdots, L\}$, it draws a direct child topic node under $c_{l-1}$ with the following distribution:
\begin{equation}
\label{eq-ncrp}
    p\left(c_{l}=k|c_{l-1}\right)=
    \left\{\begin{array}{ll}
    \frac{M_{b,k}+1}{\gamma+M_{b,c_{l-1}}+\lvert \Lambda(c_{l-1})\rvert } & \text { if } k\text { is an existing child node of $c_{l-1}$},\\
    \frac{\gamma}{\gamma+M_{b,c_{l-1}}+\lvert \Lambda(c_{l-1})\rvert} & \text { if } k\text { is a new child node},
    \end{array}\right.
\end{equation}
where $M_{b,k}$ is the number of previous document visits at topic node $k$ in the corpus of brand $b$, specially having $M_{b,c_1}=i-1$ in the $i$th trial. The hyperparameter $\gamma$ governs the likelihood of creating a new child node in topic hierarchy. Laplace smoothing is additionally incorporated to prevent zero probability values, where $|\Lambda(c_{l-1})|$ represents the number of existing child nodes descending from $c_{l-1}$.
The nCRP allows continuous change and branching of topic groups in a hierarchy, where the total number of topics is decided by data.

Each review is allowed to have multiple paths through the tree as there could be multiple semantic themes in the same review. Specifically, each review sentence is assigned to a single path with $L$-level topics based on the brand-specific nCRP prior.
Once a path $\boldsymbol{c}_{d,s}$ is selected for a review sentence indexed by $(d,s)$, the generative process of words in the sentence reduces to a flat LDA model \citep{lda}. In this case, sentence words $\{w_{d,s,n}\}_{n=1}^{N_{d,s}}$ are respectively assigned to different levels $\{l_{d,s,n}\}_{n=1}^{N_{d,s}}$ along the selected path based on an $L$-dimension multinomial distribution $\text{Multi}(\boldsymbol{\theta}_{d,s})$. $\boldsymbol{\theta}_{d,s}$ is drawn from a symmetric Dirichlet prior on level distributions: $\boldsymbol{\theta}_{d,s}\sim \text{Dir}(\boldsymbol{\alpha})$, where $\boldsymbol{\alpha}=(\alpha,\cdots,\alpha)^T$.
The review sentence path $\boldsymbol{c}_{d,s}$ and each word's level assignment $l_{d,s,n}$ jointly specifies the corresponding topic assignment of word: $k_{d,s,n} \equiv c_{d,s,l_{d,s,n}}$.
After that each sentence word is drawn from the word distribution $\boldsymbol{\phi}_{k_{d,s,n}}$ independently conditioned on its topic assignment. 

We consider the polarity score $y_d$ accompanying each review $d$ as a response variable in the regression, which is referred to as polarity response hereafter. It relies on brand-specific regression parameters $\boldsymbol{\beta}^b$ corresponding to the extracted topics $\{k\}$, and uses the empirical distribution of review $d$ over topics as regressors: ${x}^d_{k}=\frac{1}{N_{d}} \sum_{s=1}^{S_d} \sum_{n=1}^{N_{d, s}}  \mathbb{I}\left(k_{d, s, n}=k \right)$. 
The fitting of polarity response $y_d$, in turn, helps to inform the recovery of topics by incorporating their association with extremes on these topics.
Specifically, the brand-specific topic regression parameters $\boldsymbol{\beta}^b$ indicate customers' sentiment polarities towards various topics or aspects of each brand, which can be used for ranking brand performances in multiple aspects.

\begin{figure}[htb]
\centering
\includegraphics[scale=0.3]{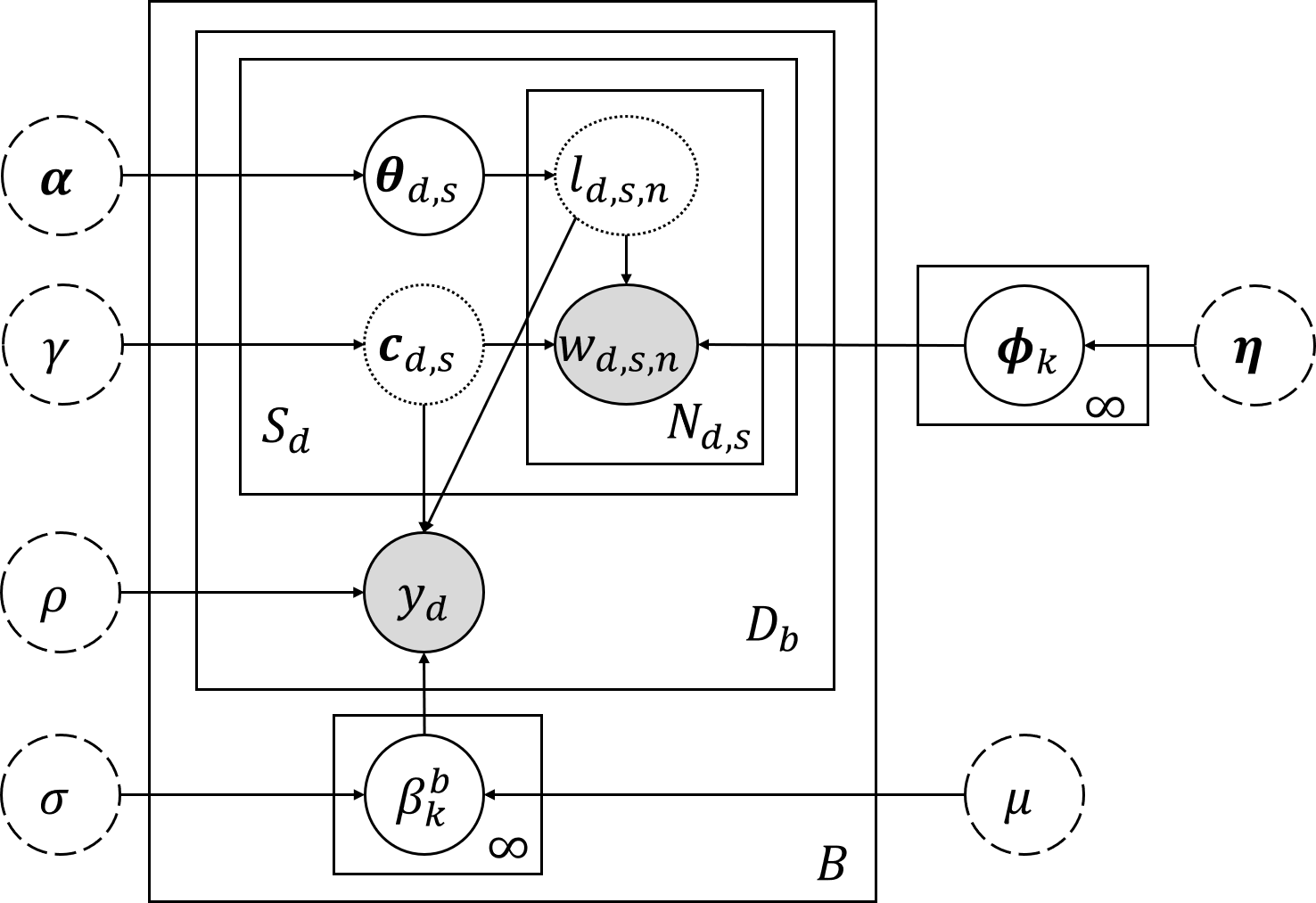}
\caption{Graphical representation of MH-STM. The shaded circles represent observable variables, dashed circles represent hyperparameters, dotted circles represent latent variables, and the rest solid circles represent model parameters.}
\label{fig-method-1}
\end{figure}

A generative procedure summarizing the proposed MH-STM is illustrated in Algorithm \ref{alg0}. Figure~\ref{fig-method-1} shows a plate notation of the generative procedure accordingly.
It is assumed that only review words $\{w_{d,s,n}\}$ and overall polarity responses $\{y_d\}$ are observed, and they are both affected by an infinite number of tree-structured latent topics that are positioned by corresponding assignments of path $\boldsymbol{c}_{d,s}\sim  \text{nCRP}(\gamma, b)$ and level $l_{d,s,n}\sim \text{Multi}(\boldsymbol{\theta}_{d,s})$ in the tree.
The observed review words are generated by multinomial word distributions $\boldsymbol{\phi}_k$ conditioned on their topic assignments, and review's polarity response variables are explained by the topic regression coefficients $\beta^b_k$ specific to each brand.  

\begin{algorithm}[ht]
\begin{spacing}{1.2}
    \caption{Generative procedure of reviews in MH-STM}
    \label{alg0} 
    \begin{itemize}[leftmargin=0pt,topsep=0pt]
    \item For each topic node $k$ in the tree:
    \begin{enumerate}[leftmargin=15pt,topsep=0pt]
        \item Draw a multinomial word probability  $\boldsymbol{\phi}_k\sim \text{Dir}(\boldsymbol{\eta})$ specific to the topic $k$.
        \item For each brand $b\in \{1,\cdots,B\}$:
        \begin{enumerate}
            \item Draw a brand-specific regression parameter $\beta^b_k \sim \mathcal{N}(\mu,\sigma^2)$.
        \end{enumerate}
    \end{enumerate}
    
    \item For each review  $d\in\{1,\cdots,D_b\}$ of brand $b$:
    \begin{enumerate}[leftmargin=15pt,topsep = 0 pt]
        \item For each sentence $s \in \{1,\cdots,S_d\}$ of review $d$:
        \begin{enumerate}
            \item Draw a path $\boldsymbol{c}_{d,s}\sim  \text{nCRP}(\gamma, b)$, where $\boldsymbol{c}_{d,s}=(c_{d,s,1},\cdots,c_{d,s,L})^T$ includes each level's topic along the path.
            \item Draw an $L$-dimensional topic level multinomial distribution $\boldsymbol{\theta}_{d,s} \sim \text{Dir}(\boldsymbol{\alpha})$.
            \item For each word token $n \in \{1,\cdots,N_{d,s}\}$ in the review-sentence pair $(d,s)$:
            \begin{enumerate}
                \item Draw topic level $l_{d,s,n} \sim \text{Multi}(\boldsymbol{\theta}_{d,s})$, which specifies the corresponding topic $k_{d,s,n} \equiv c_{d,s,l_{d,s,n}}$ along the selected path.
                \item Draw word $w_{d,s,n}\sim \text{Multi}(\boldsymbol{\phi}_{k_{d,s,n}})$.
            \end{enumerate}
        \end{enumerate}
        \item Draw polarity response $y_d \sim \mathcal{N}( (\boldsymbol{x}^d)^T \boldsymbol{\beta}^b,\rho^2)$, where  ${x}^d_{k}=\frac{1}{N_{d}} \sum_{s=1}^{S_d} \sum_{n=1}^{N_{d, s}}  \mathbb{I}\left(k_{d, s, n}=k \right)$. 
    \end{enumerate}
    \end{itemize}
\end{spacing}
\end{algorithm}  

\subsection{Enhancing topic-word association via P\'olya urn scheme}
\label{method-2}
As illustrated in Figure \ref{fig-method-1}, the multinomial word distribution $\boldsymbol{\phi}_k$ conditioned on topic $k$ is usually drawn from a symmetric Dirichlet prior: $\text{Dir}(\boldsymbol{\eta})$, where $\boldsymbol{\eta}=(\eta,\ldots,\eta)^T$.
In practice, however, it is common to deal directly with the ``collapsed'' distribution by integrating over the topic-word multinomial $\boldsymbol{\phi}_k$ \citep{mimno2011optimizing}:
\begin{equation}
\label{eq-HPU-1}
    p(w\mid k,\eta,\mathcal{W}) = \frac{N_{k,w}+\eta}{\sum_{v=1}^{V}\left(N_{k,v}+\eta\right)},
\end{equation}
which is a function of the hyperparameter $\eta$ and the number
of each word assigned to that topic, i.e., $N_{k,v}$.
This collapsed distribution is known as Dirichlet compound multinomial (DCM) or multivariate P\'olya distribution \citep{doyle2009accounting}.
It breaks the assumption of conditional
independence between word tokens given topics, but has been proven to show substantially better performance than traditional multinomials in the modeling of text documents \citep{wordburstiness}. In the following parts, we start with introducing the existing simple P\'olya urn scheme that constructs DCM in flat topic models, and then we propose a novel hierarchical P\'olya urn scheme that adapts to topic hierarchy.


\subsubsection{Simple P\'olya urn scheme}
The process for generating a sequence of words following the DCM distribution is known as the simple P\'olya urn (SPU) scheme \citep{mahmoud2008polya}, which employs an analogy of urns with colored balls to interpret the topic-word distribution. 
Let $\boldsymbol{u}_k^0=\boldsymbol{\eta}$ be an urn initially containing an equal number of $\eta$ standard balls in one
of $V$ distinct colors, and $\mathbf{U}^0=[\boldsymbol{u}_1^0,\boldsymbol{u}_2^0,\cdots,\boldsymbol{u}_K^0]^T$ represent $K$ urns in the same initial states. 
In this context, the urns and different colored balls represent $K$ topics and $V$ distinct words, respectively.
Starting at time $i = 0$, each time a colored ball is randomly drawn from an urn, it is returned to the urn along with another ball of the same color.
If $\{(k_i,v_i)\}$ is a sequence of balls each in color of $v_i\in\{1,\cdots,V\}$ and drawn from the urn $k_i\in\{1,\cdots,K\}$, then the state of urns can be describe as a recurrent process as:
\begin{equation}
\label{eq-HPU-2}
    \mathbf{U}^{i+1} = \mathbf{U}^{i} + \mathbf{E}^{i} ,
\end{equation}
where $\mathbf{E}^{i}$ is a $K\times V$ matrix with $1$ in dimension of $(k_i,v_i)$ and $0$ elsewhere. 
This process is equivalent to the marginal distribution in Equation (\ref{eq-HPU-1}) with a Dirichlet prior and a multinomial likelihood, 
where the proportion of colored balls in an urn reflects the word distribution under a topic.
As a sampled color is more likely to be sampled from the same urn again in the next round, the SPU scheme enables to capture the word burstiness in a document, i.e., if a word appears once, it is more likely to appear again \citep{doyle2009accounting}.

\subsubsection{Hierarchical P\'olya urn scheme}
The SPU scheme is designed for flat topic structure, where all word tokens are equally bursty under each of topics while ignoring the hierarchical relationship among them.
\citet{Liang2023} has developed an enhanced P\'olya urn scheme that adapts the sampling weights of balls to their general-to-specific characteristics in hierarchy. However, it still adopts the assumption of equal burstiness across hierarchy, i.e., it still follows the stochastic process in Equation (\ref{eq-HPU-2}).  
Therefore, we introduce a hierarchical P\'olya urn (HPU) scheme in this part for incorporating adaptive word burstiness in hierarchical topic modeling.

Formally, given a sequence of balls $\{(k_i,v_i)\}$, we propose the HPU scheme with urns' state process as 
\begin{equation}
\label{eq-HPU-3}
    \mathbf{U}^{i+1} = \mathbf{U}^{i} + \mathbf{E}^{i} + \Tilde{\mathbf{E}}^i\odot \mathbf{A}^i,
\end{equation}
where $\Tilde{\mathbf{E}}^i$ is a $K\times V$ matrix with $1$ in each dimension of $\{(k,v_i): k\in {\rm V}(k_i)\}$ and $0$ elsewhere, $\mathbf{A}^i$ is a $K \times V$ addition weight matrix that stores the weights of additional balls put to all ancestor urns in ${\rm V}(k_i)$, and $\odot$ performs Hadamard product of two matrices.
According to the process specified by Equation (\ref{eq-HPU-3}), when a colored ball is drawn from a certain urn, the selected ball and a new ball in the same color are put back to the urn, 
and additionally some weighted balls in the same color will be placed to the urn's all ancestor urns. 
By applying this HPU scheme in a hierarchical topic model, the consistency between parent and child topics is enhanced such that a child topic's frequent word is also likely to appear in its parent.

The additionally placed balls are weighted by their general-to-specific characteristics along the hierarchy.
Specifically, we define the addition weight matrix $\mathbf{A}^i = [a^i_{kv}]_{K\times V}$  as
\begin{equation}
\label{eq-method-addition-matrix}
\begin{aligned}
    a^i_{kv}&=\min{\left(\frac{\text{Ent}(\Lambda (k) \mid v)}{\text{Ent}(\Lambda (k))},1\right)},\\
    \text{Ent}(\Lambda (k)) &= -\sum_{k'\in \Lambda (k)} p(k')\log p(k'), \\
    \text{Ent}(\Lambda (k)\mid v) &= -\sum_{k'\in \Lambda (k)} p(k'\mid v)\log p(k'\mid v),
\end{aligned}
\end{equation}
where we use the index variables $k$ and $v$ for denoting the corresponding indexed urn (topic) and ball (word) respectively. $\text{Ent}(\Lambda (k) \mid v)$ and $\text{Ent}(\Lambda (k))$ represent Shannon's entropy on the set of $k$'s direct children $\Lambda (k)$ with and without the specified word $v$ respectively, which measures the uncertainties of distinguishing one child topic from another in $\Lambda(k)$.

The addition weights specified above range in $[0,1]$, and their values rely only on intermediate topic sampling results with no requirement for external knowledge. When the information gain $\text{Ent}(\Lambda (k))-\text{Ent}(\Lambda (k) \mid v)\leq 0$, it triggers the maximum addition weight $a_{kv}=1$. It is desired that a word should be more weighted in the parent node $k$ if it is shared by each child node in $\Lambda(k)$. 
In contrast, it triggers the minimum addition weight $a_{kv}=0$ when $\text{Ent}(\Lambda (k) \mid v)=0$, which means a word should be less weighted in the parent node $k$ if it is associated mainly with a specific child topic in $\Lambda(k)$.
Under this rule, we achieve a desired concept tree that places semantic notions of generality to the root and the counterpart of specificity to the leaves \citep{tumpa2018document}.

A sequence of words under topics generated by the HPU process defines a set of topic-specific word distributions as:
\begin{equation}
    p(w\mid k,\eta,\mathcal{W}) = \frac{W_{k,w}+\eta}{\sum_{v=1}^{V}\left(W_{k,v}+\eta\right)},
\end{equation}
where $W_{k,v}$ denotes the cumulative weights of each word $v$ under the topic $k$. It can be seen as a weighted version of the classical DCM in Equation (\ref{eq-HPU-1}). 

\begin{figure}[ht]
	\centering
	\setlength{\abovecaptionskip}{2pt}
	\subfigure[Simple P\'olya Urn (SPU)]{
	    \label{fig-method-2a}
	    \begin{minipage}{0.97\linewidth}
	    \centering	    
	    \setlength{\abovecaptionskip}{2pt}
	    \includegraphics[scale=0.3]{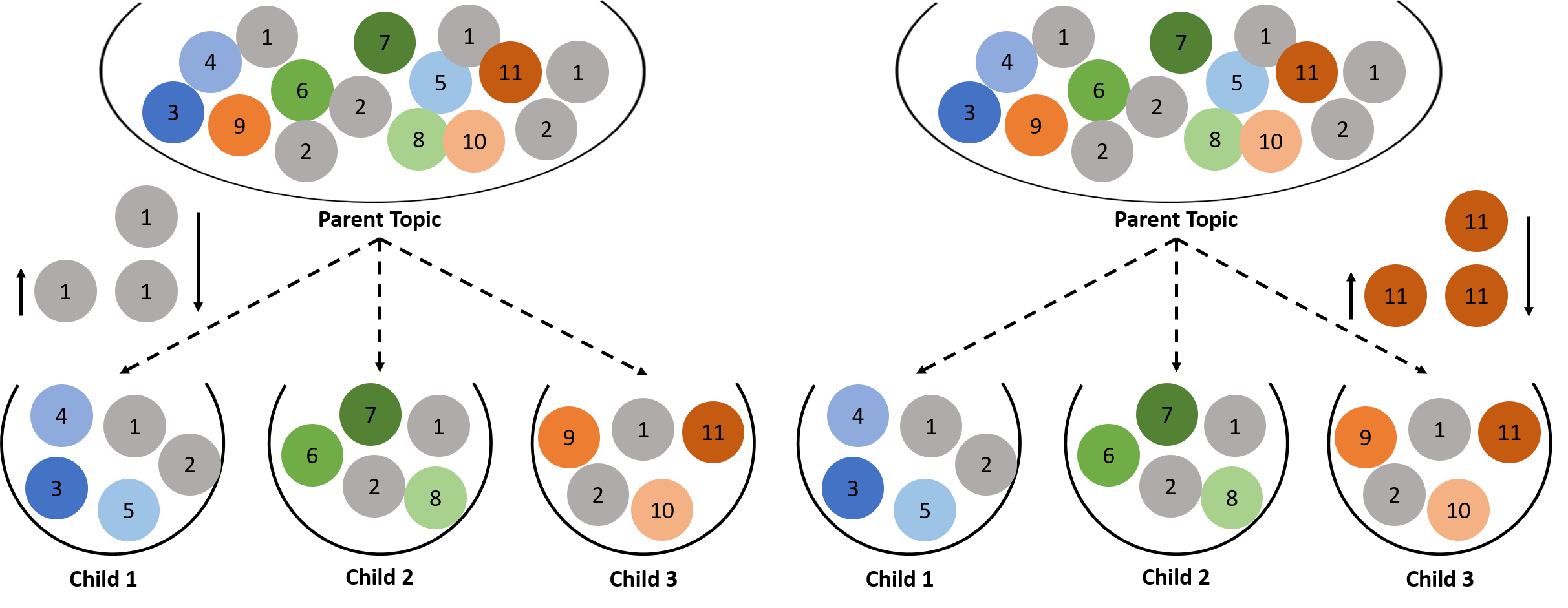}
	    \end{minipage}
	}    
	\subfigure[Hierarchical P\'olya Urn (HPU)]{
	    \label{fig-method-2b}
	    \begin{minipage}{0.97\linewidth}
	    \centering	    			\includegraphics[scale=0.3]{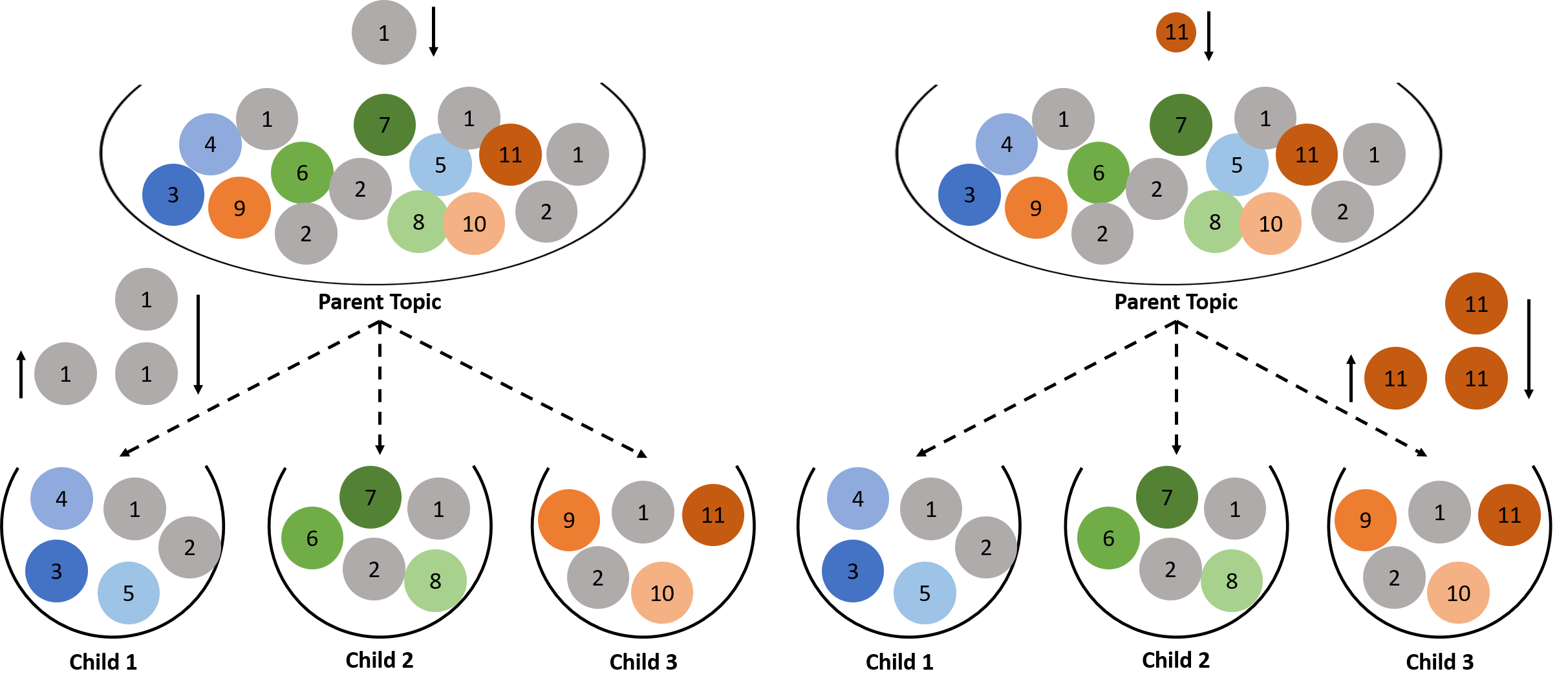}
	    \end{minipage}
	} 
	\caption{ Comparison of simple and hierarchical P\'olya urn schemes in hierarchical topic modeling. Urns represent topics and balls with different numbers/colors represent different words. Grey balls (No.1-2) represent general words that are shared among child topics, while balls in color of blue (No.3-5), green (No.6-8), and orange (No.9-11) are specialized words for each child topic, respectively. }
	\label{fig-method-2}
\end{figure}

Figure~\ref{fig-method-2} illustrates some sampling instances of the proposed HPU scheme in comparison with SPU, which are both applied to a two-level hierarchy.
Compared to SPU, HPU adds additional words to the parent topics with weights that are consistent with the general-to-specific characteristics of different words. For example, a general word $v$ shared by different child topics of $k$ leads to a lower information gain and a higher addition weight of $a_{kv}$ according to Equation (\ref{eq-method-addition-matrix}), which makes it more bursty to the parent topic $k$ (e.g., the left case in Figure~\ref{fig-method-2b}). On the contrary, a word specific to a certain child topic $k'\in \Lambda (k)$ brings higher information gain in differentiating the branches under parent topic $k$, leading to a lower value of $a_{kv}$ that prevents $v$ to be assigned to the parent topic $k$ (e.g., the right case in Figure~\ref{fig-method-2b}).

Overall, SPU assumes equal word burstiness under various topics, while HPU scheme produces word-specific burstiness that adapts to topic hierarchy, where the general words are more likely to repeat within a document around the root topic.  
Therefore, the root node tends to place probability mass on general words, and the leaf nodes place on specific words, such that the neutral topics at the top that are general to all brands are distinct from the polarised topics at the bottom that are specific to different brands, leading to accurate brand comparison in the shared topic hierarchy.

\subsection{Model inference} 
\label{method-3}
The proposed model is parameterized by $\Theta = (\boldsymbol{\theta},\boldsymbol{\phi},\boldsymbol{\beta})$.
Overall, we employ a probabilistic frame of stochastic EM that is commonly adopted for the inference of supervised topic models \citep{boyd2010,shlda}. 
The stochastic EM algorithm alternates between a Gibbs sampling E-step and an optimization M-step.  Specifically, in the E-step, we conduct a collapsed Gibbs sampling that employs the proposed HPU scheme to construct a Markov chain over two groups of latent variables, i.e., the path assignment $\boldsymbol{c}$ of each review sentence and the topic level assignment $l$ of each word. During the sampling process, one latent group in $\{\boldsymbol{c},l\}$ is sampled conditioned on the other by a set of conditional independence assumptions illustrated in Figure~\ref{fig-method-1}, and all other variables such as $\boldsymbol{\phi},\boldsymbol{\theta}$ are integrated out.
In the M-step, we optimize the multi-level topic regression parameters $\boldsymbol{\beta}$ 
by maximum likelihood estimation (MLE) given samples of $\{\boldsymbol{c},l\}$ in E-step. Let us provide a detailed explanation of the two steps below.

\textbf{Sampling $\boldsymbol{c}$ in E-step:} Conditioned on the observed words $\boldsymbol{w}_{d,s}$ and its corresponding polarity response $y_d$ of a certain review-sentence pair $(d,s)$ under brand $b$, as well as fixed topic levels $\boldsymbol{l}_{d,s}$ of the words $\boldsymbol{w}_{d,s}$, the probability of assigning $(d,s)$ to a path $\boldsymbol{c}$ is
\begin{equation}
\label{eq-sample-c-1}
\begin{aligned}
    & p(\boldsymbol{c}_{d,s}=\boldsymbol{c}|\boldsymbol{c}^{-(d,s)},\boldsymbol{l}_{d,s},\boldsymbol{w}_{d,s},y_d,b,\boldsymbol{\beta}^b) \\
    & \propto p(\boldsymbol{w}_{d,s}|\boldsymbol{c}_{d,s}=\boldsymbol{c},\boldsymbol{l}_{d,s}) \times
    p(y_d|\boldsymbol{c}_{d,s}=\boldsymbol{c},\boldsymbol{l}_{d,s},\boldsymbol{\beta}^b) \times
    p(\boldsymbol{c}_{d,s}=\boldsymbol{c}|\boldsymbol{c}^{-(d,s)},b),
\end{aligned}
\end{equation}
where $\boldsymbol{c}^{-(d,s)}$ represents the path assignments of all review sentences other than $(d,s)$ in review corpus. 
From the perspective of Bayes' rule, the first two terms can be seen as the likelihood of observed words and polarity response given a selected path, and the third term is the prior on path $\boldsymbol{c}_{d,s}$ implied by the nCRP under brand $b$.
Specifically, the prior probability of  $\boldsymbol{c}_{d,s}$ under brand $b$ is derived by the nCRP in Equation (\ref{eq-ncrp}) as
\begin{equation}
\label{eq-sample-c-2}
\begin{aligned}
    & p(\boldsymbol{c}_{d,s}=\boldsymbol{c}|\boldsymbol{c}^{-(d,s)},b) \\
    & \propto\left\{ 
    \begin{array}{ll}
    \prod_{l=2}^L \frac{M^{-(d,s)}_{b,c_{l}}+1}{\gamma+M^{-(d,s)}_{b,c_{l-1}}+|\Lambda(c_{l-1})|} & \text{ if $\boldsymbol{c}$ is an existing path},\\
    \frac{\gamma}{\gamma+M^{-(d,s)}_{b,c_{l^{*}}}+|\Lambda(c_{l^{*}})|} \prod_{l=2}^{l^{*}} \frac{M^{-(d,s)}_{b,c_{l}}+1}{\gamma+M^{-(d,s)}_{b,c_{l-1}}+|\Lambda(c_{l-1})|} & \begin{subarray}{l}\text{ if $\boldsymbol{c}$ is a new path that branches} \\ \text{under an internal node at level $l^{*}$,} \end{subarray}
    \end{array}\right.
\end{aligned}
\end{equation}
where $M^{-(d,s)}_{b,c_{l}}$ is the total number of sentence visits at the topic node $c_l$ in brand $b$'s corpus excluding the review-sentence pair $(d,s)$.
Moreover, the word likelihood of $\boldsymbol{w}_{d,s}$ is derived by the collapsed multivariate P\'olya distribution that integrates out the topic-word multinomials $\boldsymbol{\phi}$:
\begin{equation}
\label{eq-sample-c-3}
\begin{aligned}
    p(\boldsymbol{w}_{d,s}|\boldsymbol{c}_{d,s}=\boldsymbol{c},\boldsymbol{l}_{d,s}) &= 
    \prod_{n=1}^{N_{d,s}} p(w_{d,s,n}\mid k=c_{l_{d,s,n}},\boldsymbol{w}^{-(d,s)},\boldsymbol{k}^{-(d,s)})\\
    & =\prod_{n=1}^{N_{d,s}} \int_{\boldsymbol{\phi}_{k}} {\phi}_{k,w_{d,s,n}} p(\boldsymbol{\phi}_{k}\mid\boldsymbol{w}^{-(d,s)},\boldsymbol{k}^{-(d,s)})\,\mathrm{d}\boldsymbol{\phi}_{k} \\
    & = \prod_{n=1}^{N_{d,s}} \mathbb{E} (\phi_{k,w_{d,s,n}} \mid \boldsymbol{w}^{-(d,s)},\boldsymbol{k}^{-(d,s)})\\
    & = \prod_{n=1}^{N_{d,s}} \frac{W^{-(d,s)}_{k=c_{l_{d,s,n}},w_{d,s,n}}+\eta}{\sum_{v=1}^{V} \left(W^{-(d,s)}_{k=c_{l_{d,s,n}},v}  + \eta\right)}, \\
\end{aligned}
\end{equation}
where $W^{-(d,s)}_{k,v}$ represents the cumulative weights of word token $v$ under the specified topic $k$ in the entire corpus excluding review-sentence pair $(d,s)$. It is analogous to the introduced P\'olya urn process that chooses a ball in $v$th color from the urn indexed by $k$ based on the weighted proportions of existing balls in it, and $\eta$ measures the initial weights of balls at the beginning of sampling. Specifically, the cumulative weights $[W_{k, v}]_{K\times V}$ of different tokens in a set of hierarchically structured topics are updated dynamically following the proposed HPU scheme, which is elaborated in Algorithm \ref{alg1}.

Finally, the conditional density of $y_d$ is derived by a normal linear regression based on brand-associated regression parameters: 
\begin{equation}
\label{eq-sample-c-4}
    y_d \mid \boldsymbol{c}_{d,s}=\boldsymbol{c},\boldsymbol{l}_{d,s},\boldsymbol{\beta}^b \sim \mathcal{N}\left( (\boldsymbol{x}^d)^T\boldsymbol{\beta}^b,\rho^2 \right),
\end{equation} 
where each $x_k^d = \frac{1}{N_{d}}  \sum_{s'=1}^{S_d}\sum_{n'=1}^{N_{d,s'}} \mathbb{I}(k_{d,s',n'}=k) $.

\textbf{Sampling $\boldsymbol{l}$ in E-step:} Alternately, when the path $\boldsymbol{c}_{d,s}$ is selected for review-sentence pair $(d,s)$ under brand $b$, we assign each word token $w_{d,s,n}$ in $(d,s)$ to a level $l$ based on the following probability:
\begin{equation}
\label{eq-sample-z-1}
\begin{aligned}
    & p(l_{d,s,n}=l|\boldsymbol{c}_{d,s},\boldsymbol{l}_{d,s,-n},w_{d,s,n},y_d,b,\boldsymbol{\beta}^b) \\
    &\propto p(w_{d,s,n}|l_{d,s,n}=l,\boldsymbol{c}_{d,s}) \times
    p(y_d|l_{d,s,n}=l,\boldsymbol{c}_{d,s},\boldsymbol{\beta}^b) \times
    p(l_{d,s,n}=l|\boldsymbol{l}_{d,s,-n}).\\
\end{aligned}
\end{equation}
The prior of $l_{d,s,n}$ can be derived by integrating out $\boldsymbol{\theta}_{d,s}$ via Dirichlet-multinomial conjugacy:
\begin{equation}
\begin{aligned}
\label{eq-sample-z-2}
    p(l_{d,s,n}=l|\boldsymbol{l}_{d,s,-n}) 
    & =\int_{\boldsymbol{\theta}_{d,s}} p(l_{d,s,n}=l\mid\boldsymbol{\theta}_{d,s})p(\boldsymbol{\theta}_{d,s}\mid\boldsymbol{l}_{d,s,-n})\,\mathrm{d}\boldsymbol{\theta}_{d,s} \\
    & =\int_{\boldsymbol{\theta}_{d,s}} {\theta}_{d,s,l} p(\boldsymbol{\theta}_{d,s}\mid \boldsymbol{l}_{d,s,-n})\,\mathrm{d}\boldsymbol{\theta}_{d,s}  = \mathbb{E} (\theta_{d,s,l} \mid \boldsymbol{l}_{d,s,-n})\\
    &= \frac{N^{-(d,s,n)}_{d,s,l}+\alpha}{\sum_{l'=1}^{L} \left(N^{-(d,s,n)}_{d,s,l'}+\alpha\right)},
\end{aligned}
\end{equation}
where $N^{-(d,s,n)}_{d,s,l}$ represents the total number of words assigned to the $l$th level in $(d,s)$ excluding $w_{d,s,n}$.
Similarly, the likelihood of word $w_{d,s,n}$ given its level assignment with all observations excluding $(d,s,n)$ is derived as
\begin{equation}
\label{eq-sample-z-3}
p(w_{d,s,n}\mid l_{d,s,n}=l,\boldsymbol{c}_{d,s}) = 
    \frac{W^{-(d,s,n)}_{k=c_{d,s,l},w_{d,s,n}}+\eta}{\sum_{v=1}^{V} \left(W^{-(d,s,n)}_{k=c_{d,s,l},v}  + \eta\right)}.
\end{equation}
And the conditional density of $y_d$ is in the same form with Equation~(\ref{eq-sample-c-4}).

\textbf{Optimizing $\boldsymbol{\beta}$ in M-step:} 
Given empirical regressors $x_k^d = \frac{1}{N_{d}}  \sum_{s=1}^{S_d}\sum_{n=1}^{N_{d,s}} \mathbb{I}(k_{d,s,n}=k)$ from the samples of $(\boldsymbol{c},\boldsymbol{l})$ in E-step, we update the estimation of multi-level topic regression parameters $\boldsymbol{\beta}$ via MLE on observed polarity responses $\boldsymbol{y}$. That is $\hat{\boldsymbol{\beta}}=\mathop{\arg\max}\limits_{\boldsymbol{\beta}} \mathcal{L}(\boldsymbol{y},\boldsymbol{\beta})$ with 
\begin{equation}
\label{eq-optimize-beta}
\mathcal{L}(\boldsymbol{y},\boldsymbol{\beta})=-\frac{1}{2 \rho^2} \sum_{b=1}^B\sum_{d=1}^{D_b}\left(y_d-({\boldsymbol{x}}^d)^T \boldsymbol{\beta}^b\right)^2.
\end{equation}

The stopping criterion of algorithm is defined on the difference between log-likelihoods of observed review documents $\mathcal{D}=\{\boldsymbol{y},\mathcal{W}\}$ in two successive steps: $\mathcal{L}(\mathcal{D},{\Theta}^{i+1})-\mathcal{L}(\mathcal{D},{\Theta}^{i}) < \epsilon$.
The overall procedure of model inference is summarized in Algorithm \ref{alg1}. It requires four hyperparameters: $\gamma$, $\eta$, $\alpha$, $\rho$. We set the variance of polarity response as $\rho^2=0.5$ following \citet{shlda}, a symmetric Dirichlet parameter $\eta=0.1$ following \citet{hlda}, and update other hyperparameters $\gamma, \alpha$ of hierarchical topic modeling via Bayesian optimization in practice.    

In each iteration of model inference, the computation burden falls mainly on the E-step that is affected by the vocabulary size $V$ and tree depth $L$. The vocabulary size $V$ mainly influences the update of addition weight matrix $\mathbf{A}$ in E-step. As the computation complexity of $\mathbf{A}$’s update is $\mathcal{O}(KV)$, this step will scale linearly with $V$.
Moreover, the depth $L$ of topic tree mainly influences the path and level samplings in E-step. Specifically, the total computation complexity of E-step with respect to the tree depth $L$ is $\mathcal{O}(\sum_{b=1}^B\sum_{d=1}^{D_b}\sum_{s=1}^{S_d}(1+N_{d,s})L)$, and this step will scale linearly with $L$.

\begin{algorithm}[H]
\begin{spacing}{1.2}
\SetNoFillComment
\SetAlgoLined
\small
\KwIn{review collection $\{\{\boldsymbol{w}_d,y_d\}_{d=1}^{D_b}\}_{b=1}^{B}$, hyperparameters $\gamma$, $\eta$, $\alpha$, $\rho$.} 
\KwOut{Topic hierarchy with brand-associated regression parameters of topics. } 
\caption{Stochastic EM algorithm for MH-STM with HPU scheme.} 
\label{alg1}
\BlankLine
Initialize root topic of the tree and the addition weight matrix $\mathbf{A}^{0}=\mathbf{A}^{1}=[0]_{K \times V}$\;

\For{each iteration $i \in \{1,\cdots,I\}$} 
{ 
    \tcc{Sampling paths in E-step} 
    \For{each review-sentence pair $(d,s)$ under brand $b$ }
    {        

            Exclude $(d,s)$ associated with its path assignment $\boldsymbol{c}_{d,s}$ from count variables:\\
            
                \hspace{5mm} $M^{-(d,s)}_{b,c_{d,s,l}} \leftarrow M_{b,c_{d,s,l}} -1$ for $\forall l\in \{1,\dots,L\}$\;
            
                \hspace{5mm} $W^{-(d,s)}_{k_{d,s,n},w_{d,s,n}} \leftarrow W_{k_{d,s,n},w_{d,s,n}}-1$ for $\forall n\in\{1,\cdots,N_{d,s}\}$\;

                \hspace{5mm} $W^{-(d,s)}_{k,w_{d,s,n}} \leftarrow W_{k,w_{d,s,n}} - a^{i-1}_{kw_{d,s,n}} $ for $\forall n\in\{1,\cdots,N_{d,s}\}$, $\forall k \in \text{V}(k_{d,s,n})$\;
            
		Sample a new path $\boldsymbol{c}_{d,s}$ for $\boldsymbol{w}_{d,s}$ via Equations~(\ref{eq-sample-c-1})-(\ref{eq-sample-c-4})\;
		Update count variables with the new path $\boldsymbol{c}_{d,s}$:\\

                \hspace{5mm} $M_{b,c_{d,s,l}}  \leftarrow  M^{-(d,s)}_{b,c_{d,s,l}}+1$ for $\forall l\in \{1,\dots,L\}$\;

                \hspace{5mm} $W_{k_{d,s,n},w_{d,s,n} } \leftarrow  W^{-(d,s)}_{k_{d,s,n},w_{d,s,n}} +1$ for $\forall n\in\{1,\cdots,N_{d,s}\}$\;

                \hspace{5mm} $W_{k,w_{d,s,n}} \leftarrow W^{-(d,s)}_{k,w_{d,s,n}} + a^{i}_{kw_{d,s,n}} $ for $\forall n\in\{1,\cdots,N_{d,s}\}$, $\forall k \in \text{V}(k_{d,s,n})$\;

    }
    \tcc{Sampling levels in E-step}
    \For{each review-sentence pair $(d,s)$} 
    {        

          \For{each word $w_{d,s,n},n=1,\dots,N_{d,s}$}
          {
            Exclude $w_{d,s,n}$ associated with its level assignment $l_{d,s,n}$ from count variables:  \\
            \hspace{5mm} $N^{-(d,s,n)}_{d,s,l_{d,s,n}} \leftarrow N_{ d,s,l_{d,s,n}} -1 $\;  
            \hspace{5mm} $W^{-(d,s,n)}_{k_{d,s,n},w_{d,s,n} } \leftarrow W_{k_{d,s,n},w_{d,s,n}}-1$\;

            \hspace{5mm} $W^{-(d,s,n)}_{k,w_{d,s,n}} \leftarrow W_{k,w_{d,s,n}} - a^{i}_{kw_{d,s,n}} $ for $\forall k \in \text{V}(k_{d,s,n})$\;

		Sample a new level $l_{d,s,n}$ for $w_{d,s,n}$ via Equations~(\ref{eq-sample-z-1})-(\ref{eq-sample-z-3})\;
		Update count variables with the new level $l_{d,s,n}$:\\
                \hspace{5mm} $ N_{d,s,l_{d,s,n}} \leftarrow N^{-(d,s,n)}_{d,s,l_{d,s,n}} +1 $\;  
                \hspace{5mm} $ W_{k_{d,s,n},w_{d,s,n}} \leftarrow  W^{-(d,s,n)}_{k_{d,s,n},w_{d,s,n}}+1$\;

                \hspace{5mm} $W_{k,w_{d,s,n}} \leftarrow W^{-(d,s,n)}_{k,w_{d,s,n}} + a^{i}_{kw_{d,s,n}} $ for $\forall k \in \text{V}(k_{d,s,n})$\;
          }
    }
    Update addition matrix $\mathbf{A}^{i+1}$ via Equation~(\ref{eq-method-addition-matrix})\;
    \tcc{M-step}
    Update $\boldsymbol{\beta}$ by MLE on Equation~(\ref{eq-optimize-beta})\; 
} 
\end{spacing}
\end{algorithm}

\section{Simulation study}
\label{simulation}
In this section, we conduct different simulation scenarios for evaluating the performance of our proposed model extensively.
We first consider a small-sized demonstration in Section \ref{simulation-1} for visualization and qualitative analysis. Then we conduct quantitative comparison in two aspects on larger simulated corpora in Section \ref{simulation-2}: one is the accuracy of multi-aspect brand rankings, and the other is the quality of extracted topic hierarchy. 

Data was generated to reasonably resemble the observed review corpora under various hierarchical topic structures. Considering the hyperparameter $\eta$ governs the smoothing degree of each topic over vocabulary, e.g., a larger $\eta$ tends to generate a more smoothed topic-word distribution, we also demonstrate the simulation results under different values of $\eta$.
Moreover, we compare the proposed method with the following baseline models:
\begin{itemize}
    \item[M1:] SLDA \citep{slda} is a supervised LDA model that extends classical LDA with a response variable predicted by latent topics. We employed separate regressions on each brand corpus to obtain brand-associated aspect polarities. 
    The fixed number of topics in SLDA is set equal to the number of leaf nodes in the hierarchical model.
    \item[M2:] SHLDA \citep{shlda} is a supervised hierarchical LDA model. We also employed separate regressions on each brand corpus for brand comparison.
\end{itemize}

\subsection{Qualitative comparison with topic tree visualization}
\label{simulation-1}
We first illustrate the effectiveness of the proposed MH-STM model on a synthetic corpus with a small vocabulary. We generate a synthetic corpus that consists of 10 brands each having 200 documents from a 9-term vocabulary represented by a $3 \times 3$ grid. 
The observed words are generated by a three-level topic tree with 1, 3, 9 topics at the first level, second level, and third level, respectively. 
The root topic has a uniform distribution over the entire vocabulary grid.
Topics at the second level are uniformly distributed over the terms in each row of the grid.
Topics at the third level have full probability concentrated at a single term from the row of its parent.

\begin{figure}[!htb]
	\centering
	\setlength{\abovecaptionskip}{2pt}
	\subfigure[MH-STM]{
	    \label{fig-simulation-1a}
	    \begin{minipage}{0.97\linewidth}
	    \centering	    
	    \setlength{\abovecaptionskip}{2pt}
	    \includegraphics[scale=0.27]{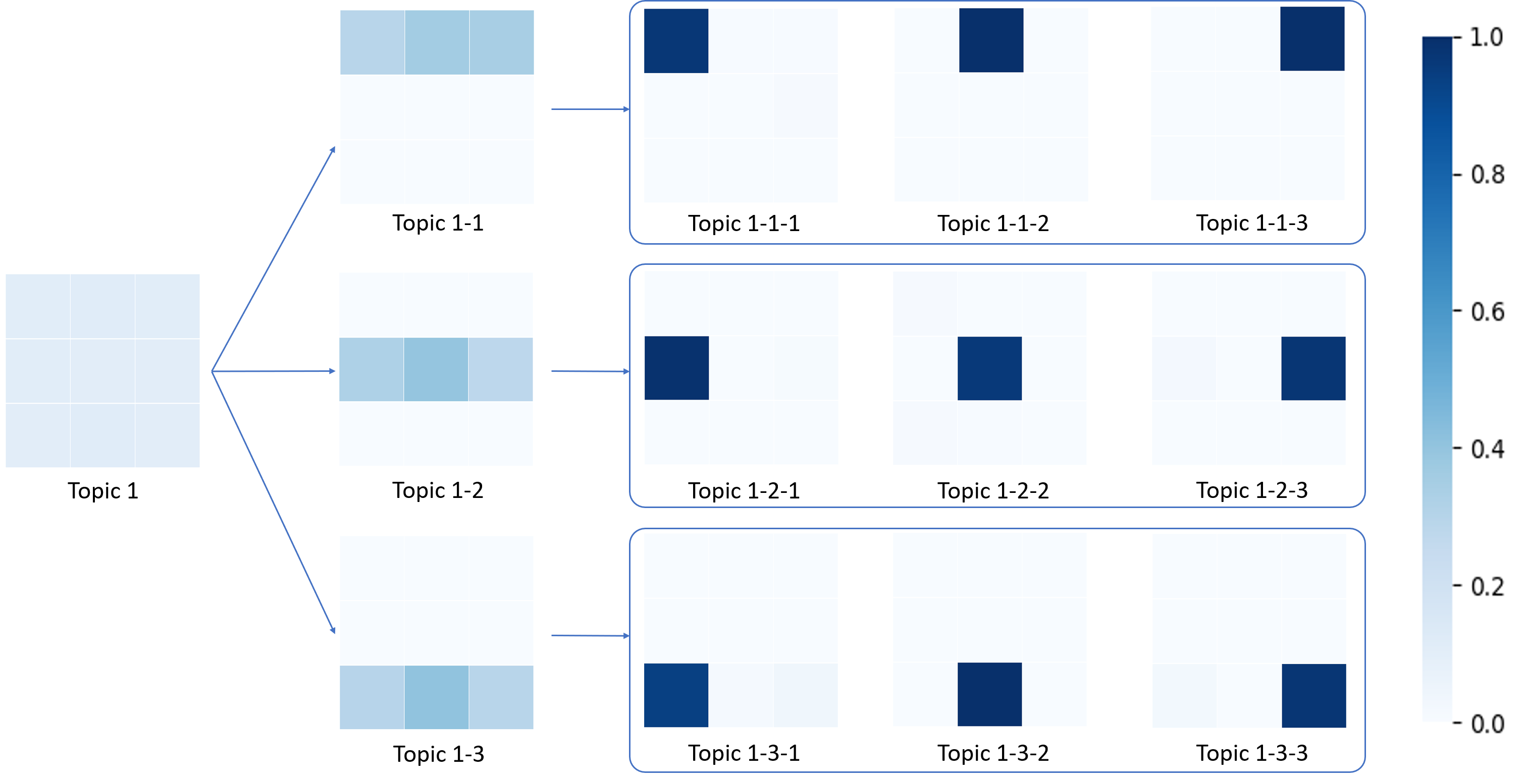}
	    \end{minipage}
	}    
	\subfigure[SHLDA]{
	    \label{fig-simulation-1b}
	    \begin{minipage}{0.97\linewidth}
	    \centering	    			\includegraphics[scale=0.27]{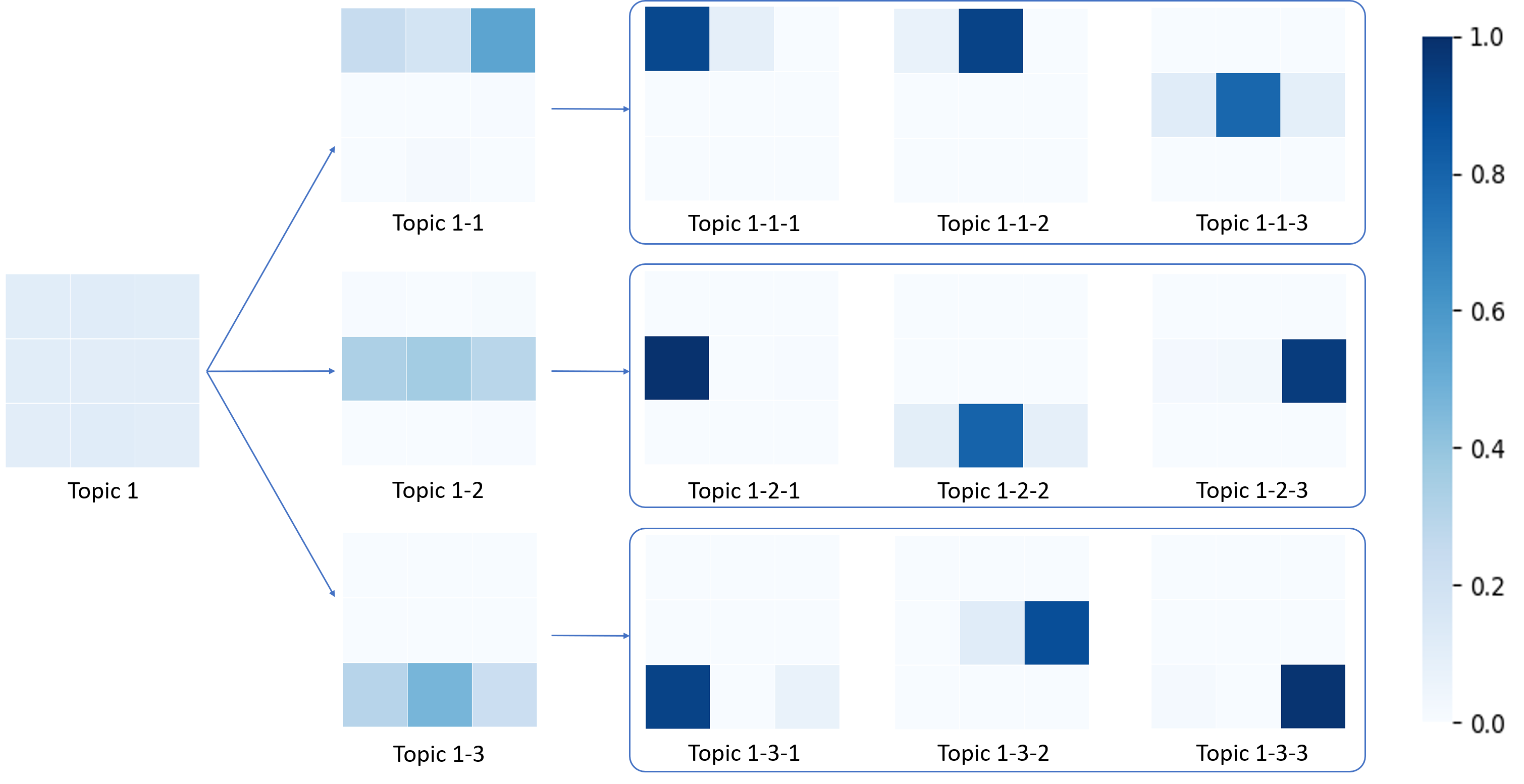}
	    \end{minipage}
	} 
	\caption{Topic trees inferred from synthetic data. Each cell corresponds to a single word, and is shaded with intensity proportional to the probability of each word in the topic.}
	\label{fig-simulation-1}
\end{figure}

The results of topic hierarchy inferred from the synthetic data above are visualized in Figure~\ref{fig-simulation-1}. The proposed MH-STM model successfully recovers the original hierarchical structure, and the inferred topics by MH-STM are almost identical to the original topics. 
By contrast, in the tree derived by SHLDA, the first and second level topics are almost identical to the original topics, while the third level topics are disturbed by some noise words. In addition, the topic tree inferred by SHLDA exhibits higher disorders in hierarchical affinity. For example in Figure~\ref{fig-simulation-1b}, the topic 1-1-3 is more close to the second-level topic 1-2 than its direct parent topic 1-1. This observation is consistent with the quantitative comparison results of hierarchical affinity scores in the following Table \ref{tab-simul-2} and Table \ref{tab-case-2}, that MH-STM is much preferred for reasonable parent-child relations in extracted hierarchy.

\subsection{Quantitative analysis}
\label{simulation-2}
To have a comprehensive examination of the proposed model, 
we evaluate the proposed method on a set of synthetic corpora that are generated under different scenarios of topic structures and distribution parameters. 
Each experimental corpus is generated from a three-level topic hierarchy with a vocabulary of 100 terms.  
We consider different probability distributions specific to totally $B=10$ brands on the topic hierarchy, and generate $D_b = 200$ synthetic reviews respectively for each brand relying on the generative process in Section~\ref{method-1}, with each review including $S_d=5$ sentences in average length of $N_{d,s}=10$ words. 

In each simulation scenario, we consider the brand topic regression parameter $\beta_k^b \sim \mathcal{N}(\mu,\sigma^2_{l(k)})$, where $\mu=0$ and $\boldsymbol{\sigma}=(\sigma_1,\sigma_2,\sigma_3)^T = (1,2,3)^T$, depending on the topic level $l(k)\in\{1,2,3\}$ in the three-level hierarchy. This formulation induces higher variance of brand polarities towards leaf topics, which is consistent with the real-world observation in Figure~\ref{fig-case-1}. We assume a symmetric Dirichlet with hyperparameter $\boldsymbol{\alpha}=(1.0,1.0,1.0)^T$ for the 3-dimensional topic proportion vector $\boldsymbol{\theta}\sim\text{Dir}(\boldsymbol{\alpha})$ of each sentence. We also set $\rho=1$ as the standard deviation for generating polarity responses in all of simulation cases.

To assess the effectiveness of the proposed model in a variety of experimental scenarios, we designed hierarchical structures following the simulation settings similar to the settings in other hierarchical topic models \citep{hlda,blei2010nested}. 
For example, the scales of their ground-truth topic hierarchies are generally 3-level tree structures, ranging 3-5 branches from the root. Similarly, we designed several 3-level topic hierarchies denoted as $n (t_{1}, \ldots, t_{n})$: 3(3,2,4), 4(6,5,2,4), 5(3,5,5,2,1). 
Here $n$ is the number of branches from the root and $t_{j}$ is the number of children under the $j$th branch. 
Given the fixed tree structure with finite paths of topics, the path assignments of each brand's reviews are decided by brand-specific distributions over paths that are randomly generated.
Moreover, we consider different smoothing parameters $\eta\in\{0.01,0.1,1.0\}$ for drawing the topic components.

\subsubsection{Multi-aspect brand ranking evaluations} 

We report the multi-aspect brand ranking results based on several evaluation metrics that are commonly used in ranking tasks: (1) Two correlation coefficients respectively for Spearman’s correlations and Kendall’s Tau. Both metrics measure how well the predicted brand aspect rankings correlate with their ground-truth aspect ratings under each leaf topic.
(2) Average precision (AP@K) that evaluates the model’s top-K recommendation performance in ranking different brands for a specified aspect. We treat each leaf topic as a query, and the top K of brands ranked by ground-truth aspect ratings as the relevant items. The AP@K measures how many of the top-K brands recommended by a model are included in the relevant items and how highly they are ranked.
A higher AP@K indicates better top-K recommendation performance of a model. Specifically, we report the average results of AP@5 corresponding to top-5 recommendation for each leaf topic.

For each model, we repeat the inference algorithm 20 times with various initializations driven by different random seeds and take the average results.
Table~\ref{tab-simul-1} summarizes the average results of different metrics as well as their standard deviations for each model. Overall, our proposed MH-STM model achieves the best performance in multi-aspect brand comparison. Although the performance decays with larger values of $\eta$, MH-STM still dominates other alternatives. The performance decay under large $\eta$ is expected, as large values of $\eta$ produce highly smoothed word distributions across topics, leading to difficulty in separating one topic from another based on their word co-occurrence rules.

\begin{table}[!ht]
\vspace{-10pt} 
\begin{spacing}{1.2}
\caption{Comparison results of multi-aspect brand rankings on simulated corpora.}
\centering
\label{tab-simul-1}
\begin{tabular}{llllll}
\hline
Hierarchy                    & $\eta$                   & Metrics       & SLDA & SHLDA         & MH-STM             \\ 
\hline
\multirow{9}{*}{3(3,2,4)}     & \multirow{3}{*}{0.01} & Spearman's    &  0.602 (0.013)   & 0.560 (0.021) & \textbf{0.680 (0.016)} \\
    &   & Kendall's tau &  0.511 (0.011)    & 0.492 (0.018) & \textbf{0.580 (0.014)} \\
    &   & AP@5 &  0.760 (0.009)   & 0.781 (0.012) & \textbf{0.842 (0.008)} \\
    & \multirow{3}{*}{0.1}  & Spearman's    &   0.584 (0.015)  & 0.626 (0.022) & \textbf{0.732 (0.016)} \\
    &   & Kendall's tau &   0.477 (0.013)  & 0.540 (0.018) & \textbf{0.628 (0.014)} \\
    &  & AP@5  &   0.745 (0.013)  & 0.828 (0.011) & \textbf{0.887 (0.011)} \\
    & \multirow{3}{*}{1}    & Spearman's    &   0.234 (0.022)   & 0.246 (0.024) & \textbf{0.406 (0.015)} \\
    &     & Kendall's tau &   0.180 (0.018)   & 0.190 (0.018) & \textbf{0.309 (0.012)} \\
    &   & AP@5  &   0.517 (0.016)   & 0.602 (0.019) & \textbf{0.698 (0.011)} \\ 
\hline
\multirow{9}{*}{4(6,5,2,4)}   & \multirow{3}{*}{0.01} & Spearman's    &   0.573 (0.012)   & 0.575 (0.014) & \textbf{0.640 (0.015)} \\
    &   & Kendall's tau &   0.444 (0.010)   & 0.461 (0.012) & \textbf{0.517 (0.013)} \\
    &    & AP@5   &   0.783 (0.008)  & 0.766 (0.010) & \textbf{0.802 (0.012)} \\
    & \multirow{3}{*}{0.1}  & Spearman's    &   0.479 (0.012)  & 0.563 (0.015) & \textbf{0.667 (0.015)} \\
    &   & Kendall's tau &   0.361 (0.010)   & 0.457 (0.012) & \textbf{0.542 (0.012)} \\
    &   & AP@5  &   0.676 (0.009)   & 0.777 (0.012) & \textbf{0.847 (0.008)} \\
    & \multirow{3}{*}{1}    & Spearman's    &   0.104 (0.016)   & 0.146 (0.017) & \textbf{0.214 (0.015)} \\
    &  & Kendall's tau &   0.090 (0.011)   & 0.106 (0.014) & \textbf{0.172 (0.009)} \\
    &   & AP@5   &   0.524 (0.011)   & 0.553 (0.008) & \textbf{0.612 (0.008)} \\ 
\hline
\multirow{9}{*}{5(3,5,5,2,1)} & \multirow{3}{*}{0.01} & Spearman's    &  0.529 (0.011)   & 0.505 (0.018) & \textbf{0.591 (0.011)} \\
    &    & Kendall's tau &   0.419 (0.010)   & 0.402 (0.014) & \textbf{0.467 (0.011)} \\
    &     & AP@5   &   0.769 (0.008)   & 0.783 (0.010) & \textbf{0.825 (0.010)} \\
    & \multirow{3}{*}{0.1}  & Spearman's    &   0.497 (0.011)   & 0.530 (0.009) & \textbf{0.642 (0.011)} \\
    &    & Kendall's tau &   0.378 (0.010)   & 0.422 (0.008) & \textbf{0.528 (0.009)} \\
    &      & AP@5   &   0.730 (0.008)   & 0.776 (0.009) & \textbf{0.837 (0.007)} \\
    & \multirow{3}{*}{1}    & Spearman's    &   0.087 (0.017)  & 0.196 (0.020) & \textbf{0.235 (0.013)} \\
    &   & Kendall's tau &   0.063 (0.013)   & 0.147 (0.016) & \textbf{0.176 (0.010)} \\
    &   & AP@5  &   0.538 (0.016)
   & 0.619 (0.010) & \textbf{0.642 (0.006)} \\ 
\hline
\end{tabular}
\end{spacing}
\end{table}

\subsubsection{Recovery of topic structure}
We quantitatively evaluate the performance of different methods in topic recovery by four measures, including: (1) Topic accuracy that quantifies the proportion of correct topic assignments in synthetic documents. Considering the issue of label switching between the learned topics and the ground-truth ones, we find the optimal alignment between the learned and the ground-truth topics by
Kuhn-Munkres algorithm \citep{lovasz2009matching}. 
(2) Held-out likelihood that evaluates how well the trained model explains the held-out data. It is defined as the average log-likelihood per word of a held-out corpus given the trained model, which is estimated by the importance sampling methods in \citet{wallach2009evaluation}. (3) The coherence scores of top-5 words under each topic. A higher coherence score indicates a superior topic that better captures the word co-occurrence rules in corpus. We adopt the topic coherence metric by \citet{mimno2011optimizing}: 
{\setlength\abovedisplayskip{0.2cm}
\setlength\belowdisplayskip{0.2cm}
$$
Coherence\left(k; V^{(k)}\right)=\sum_{i=2}^{5} \sum_{j=1}^{i-1} \log \frac{DF\left(v_i^{(k)}, v_j^{(k)}\right)+1}{DF\left(v_j^{(k)}\right)},
$$
}where $V^{(k)} = \left(v^{(k)}_1 , \cdots , v^{(k)}_{5} \right)$ represents the top 5 words assigned with topic $k$, $DF\left(v^{(k)}_j\right)$ is the document frequency of word $v^{(k)}_j$, and $DF\left(v^{(k)}_i , v^{(k)}_j\right)$ is the co-document frequency of both words. 
(4) The hierarchical affinity score that measures the goodness of a topic hierarchy \citep{kim2013hierarchical}. It is defined as the ratio of the average cosine similarity between the second-level topics and their direct children topics to the average cosine similarity between the second-level topics and their non-children topics at the third level. A lower hierarchical affinity score, especially lower than 1, indicates higher disorders in tree's parent-child affinity that a parent topic node tends to violate its direct children in semantic consistency.

\begin{table}[htpb]
\begin{spacing}{1.2}
\caption{Comparison results of topic quality on simulated corpora.}
\centering
\label{tab-simul-2}
\begin{tabular}{cccccc}
\hline
\multicolumn{1}{l}{Hierarchy}   & \multicolumn{1}{l}{$\eta$} & Metrics                 & SLDA & SHLDA          & MH-STM              \\ \hline
\multirow{12}{*}{3(3,2,4)}     & \multirow{4}{*}{0.01}   & Topic accuracy          &   0.566 (0.006)   & 0.546 (0.010)  & \textbf{0.583 (0.010)}  \\
                                &                         & Held-out likelihood     &   -2.713 (0.002)	& \textbf{-2.654 (0.003)}	& -2.663 (0.004)
          \\
                                &                         & Coherence               &   -1.456 (0.021)   & -1.608 (0.072) & \textbf{-1.265 (0.020)} \\
                                &                         & Hierarchical   Affinity &   ---   & 2.748 (0.486)  & \textbf{3.185 (0.298)}  \\
                                & \multirow{4}{*}{0.1}    & Topic accuracy          &   0.438 (0.004)   & 0.475 (0.009)  & \textbf{0.513 (0.008)}  \\
                                &                         & Held-out likelihood     &  -3.924 (0.003) &	\textbf{-3.881 (0.002)}	& -3.902 (0.003)
          \\
                                &                         & Coherence               &   -1.772 (0.009)   & -1.737 (0.010) & \textbf{-1.700 (0.017)} \\
                                &                         & Hierarchical   Affinity &   ---   & 1.654 (0.155)  & \textbf{1.994 (0.081)}  \\
                                & \multirow{4}{*}{1}      & Topic accuracy          &   0.146 (0.001)   & 0.192 (0.003)  & \textbf{0.225 (0.004)}  \\
                                &                         & Held-out likelihood     &   -4.520 (0.001)	& \textbf{-4.489 (0.001)} & -4.503 (0.001)
          \\
                                &                         & Coherence               &  -2.165 (0.009)  & -2.170 (0.006) & \textbf{-2.144 (0.006)} \\
                                &                         & Hierarchical   Affinity &   ---   & 1.015 (0.023)  & \textbf{1.183 (0.010)}  \\ \hline
\multirow{12}{*}{4(6,5,2,4)}   & \multirow{4}{*}{0.01}   & Topic accuracy   &   0.507 (0.007)  & 0.472 (0.006)  & \textbf{0.546 (0.007)}  \\
                                &                         & Held-out likelihood     &   -2.952 (0.003)	& \textbf{-2.869 (0.003)}	& -2.899 (0.003)
      \\
                                &                         & Coherence               &  -2.147 (0.05)  & -1.872 (0.069) & \textbf{-1.614 (0.028)} \\
                                &                         & Hierarchical   Affinity &   ---   & 1.847 (0.192)  & \textbf{3.032 (0.177)}  \\
                                & \multirow{4}{*}{0.1}    & Topic accuracy          &  0.393 (0.005)  & 0.396 (0.005)  & \textbf{0.427 (0.005)}  \\
                                &                         & Held-out likelihood     &  -4.045 (0.003)	& \textbf{-3.985 (0.002)}	& -4.013 (0.002)
         \\
                                &                         & Coherence               &   -1.994 (0.008)   & -1.984 (0.009) & \textbf{-1.868 (0.010)} \\
                                &                         & Hierarchical   Affinity &   ---   & 1.249 (0.059)  & \textbf{1.926 (0.062)}  \\
                                & \multirow{4}{*}{1}      & Topic accuracy          &   0.082 (0.001)   & 0.117 (0.001)  & \textbf{0.127 (0.001)}  \\
                                &                         & Held-out likelihood     &   -4.544 (0.000)	& \textbf{-4.490 (0.001)}	& -4.502 (0.001)
   \\
                                &                         & Coherence               &   -2.275 (0.005) & -2.286 (0.004) & \textbf{-2.241 (0.004)} \\
                                &                         & Hierarchical   Affinity &   ---   & 0.997 (0.006)  & \textbf{1.233 (0.034)}  \\ \hline
\multirow{12}{*}{5(3,5,5,2,1)} & \multirow{4}{*}{0.01}   & Topic accuracy          &   0.495 (0.007)   & 0.493 (0.010)  & \textbf{0.548 (0.008)}  \\
                                &                         & Held-out likelihood     &   -2.932 (0.002)	& \textbf{-2.815 (0.004)}	& -2.864 (0.004)
          \\
                                &                         & Coherence               &  -2.063 (0.048) & -1.832 (0.057) & \textbf{-1.634 (0.041)} \\
                                &                         & Hierarchical   Affinity &  ---  & 2.400 (0.192)  & \textbf{3.778 (0.210)}  \\
                                & \multirow{4}{*}{0.1}    & Topic accuracy          &   0.407 (0.006)   & 0.411 (0.006)  & \textbf{0.459 (0.006)}  \\
                                &                         & Held-out likelihood     &   -4.009 (0.002)	& \textbf{-3.933 (0.002)}	& -3.968 (0.002)
          \\
                                &                         & Coherence               &   -1.957 (0.009)   & -1.950 (0.009) & \textbf{-1.849 (0.008)} \\
                                &                         & Hierarchical   Affinity &   ---   & 1.425 (0.073)  & \textbf{2.488 (0.091)}  \\
                                & \multirow{4}{*}{1}      & Topic accuracy          &  0.087 (0.001) & 0.113 (0.001)  & \textbf{0.132 (0.001)}  \\
                                &                         & Held-out likelihood     &  -4.544 (0.000)	& \textbf{-4.487 (0.001)}	& -4.501 (0.001)
         \\
                                &                         & Coherence               &   -2.264 (0.006)   & -2.260 (0.003) & \textbf{-2.239 (0.004)} \\
                                &                         & Hierarchical   Affinity &   ---   & 0.988 (0.007)  & \textbf{1.251 (0.021)}  \\ \hline
\end{tabular}
\end{spacing}
\end{table}

The average values and standard deviations of different performance metrics for alternative models are presented in Table~\ref{tab-simul-2}. Overall, our proposed MH-STM model achieves the highest topic accuracy and coherence, indicating its superiority in the unsupervised discovery of latent topics.  
In comparison with SHLDA, MH-STM  produces uniformly higher hierarchical affinity scores, which demonstrates that the discovered topic trees by MH-STM show higher consistency among parent-child relations, and this result is also in accord with the qualitative comparison on visualized hierarchies in Section~\ref{simulation-1}. 
Likewise, the affinity performance degrades under a large value of smoothing parameter $\eta$, as large values of $\eta$ produce highly smoothed word distributions across topics, leading to difficulty in separating one topic from another and higher disorders in parent-child relations.
In terms of held-out likelihood, the hierarchical models (i.e., SHLDA and MH-STM) produce generally better results than the flat model of SLDA. However, the proposed MH-STM is overtaken by SHLDA in likelihood comparison. 
This loss is induced by the enhanced sampling process of HPU on the proposed model, during which the observed words are sampled based on not only their full probabilities but also their general-to-specific characteristics, such that the final results deviate slightly from a maximum likelihood. However, a slight loss of likelihood is worth for more interpretable topics and higher predictive accuracy in brand ranking. 
Moreover, the loss of held-out likelihood by MH-STM is consistent with the observations in previous studies on topic comparison \citep{chang2009reading,mimno2011optimizing} that higher likelihood of held-out documents could violate the judgement of topic coherence in an interpretable latent space.

\section{Real-data case studies}
\label{case}

In this section, we evaluate the proposed method under two representative review corpora in the real case studies: laptop review corpus and beer review corpus.

\textbf{Laptop review corpus (Data 1):} This corpus \citep{ni2019justifying} consists of customer reviews with corresponding overall ratings on Amazon that belong to four laptop brands: Dell, HP, Lenovo, and Apple. After preprocessing, it contains 100,931 reviews with a vocabulary of 1,642 unique terms.
    
\textbf{Beer review corpus (Data 2):} The beer review corpus \citep{mcauley2012learning} consists of beer reviews on RateBeer that belong to 26 brewers each regarded as a brand. Specifically, each review is accompanied by both an overall rating and four aspect-specific ratings that correspond to four performance aspects of beer: appearance, aroma, palate, and taste. Given the four aspects and their aspect-specific ratings as the ground truth, the multi-aspect brand ranking results can be validated. After preprocessing, the corpus contains 89,309 reviews with a vocabulary of 1,318 terms.

For both corpora, we implemented a series of pre-processing procedures. 
The data pre-processing on review texts includes sentence and word tokenizations, removing terms such as punctuations, digits, stop words, infrequent words, and stemming each word to its root for a dense vocabulary. As the overall ratings and various aspect ratings originally range in different scales, we map them uniformly to $[0,1]$ by performing a min-max normalization.

All models are implemented on individual sentences of reviews for leveraging the sentence-level information. 
Following the practice in \citet{shlda}, we set regression parameters of the root node to zero for both hierarchical models, i.e., SHLDA and MH-STM. 
It is reasonable to assume that root node would not change the response variable since it is associated with every document. 

\subsection{Multi-aspect brand rankings}
As the laptop review corpus does not provide ground truth aspect-specific ratings, 
we examine the multi-aspect brand ranking results only on the beer review corpus (Data 2).
Specifically, all models were trained on the entire beer corpus (Data 2) with only overall ratings, and the aspect-specific ratings were held out for validation. To align the topics extracted freely from corpus to the four specified aspects, we adopt four hold-out corpora each composed of review sentences annotated with corresponding aspect labels. Each annotated corpus is matched with the model's extracted topics and represented as a topic distribution specific to that aspect. 
After that, the brand rankings in an aspect are predicted by their mean polarity responses with respect to the aspect's distribution over topics.

We report the multi-aspect brand ranking results of different models in Table~\ref{tab-case-1}. 
Similarly, the model performance in multi-aspect brand ranking is quantitatively evaluated by the following metrics: 
(1) The average results of Spearman’s correlations and Kendall’s Tau for ranking different brands in each aspect.
(2) Mean average precision (AP@5 and AP@10) that evaluates the model’s top-5 and top-10 recommendation performances in each aspect. 
It can be seen from Table~\ref{tab-case-1} that MH-STM achieves the best performance in aspect-specific brand rankings. Overall, hierarchical models (i.e., SHLDA and MH-STM) outperform the flat model (i.e., SLDA) in producing accurate brand rankings for a specified aspect. Compared to SHLDA, MH-STM can provide an improved brand differentiation by integrating brand preference on topic distributions and an enhanced HPU scheme.

\begin{table}[ht]
\begin{spacing}{1.2}
\centering
\caption{Multi-aspect brand ranking evaluations on beer review corpus}
\label{tab-case-1}
\begin{tabular}{ccccc}
\hline
Beer aspect & Metrics   & SLDA   & SHLDA   & MH-STM \\ 
\hline
    & Spearman's & 0.246  & 0.458  & \textbf{0.586} \\
    & Kendall's tau  & 0.182    & 0.323  & \textbf{0.434} \\
    & AP@10 & 0.438  & 0.642       & \textbf{0.674} \\
\multirow{-4}{*}{Appearance} & AP@5  & 0.357  & 0.457 & \textbf{0.547} \\ 
\hline
    & Spearman's & 0.636  & 0.815  & \textbf{0.875}  \\
    & Kendall's tau  & 0.471  & 0.637   & \textbf{0.723}  \\
    & AP@10 & 0.602 & 0.897 & \textbf{0.918} \\
\multirow{-4}{*}{Aroma}  & AP@5  & 0.543   & \textbf{0.910}  & 
 \textbf{0.910}         \\ 
\hline
    & Spearman's  & 0.385 & 0.855 & \textbf{0.936}   \\
    & Kendall's tau & 0.268  & 0.673  & \textbf{0.785}   \\
    & AP@10 & 0.534 & 0.969 & \textbf{0.990} \\
\multirow{-4}{*}{Palate} & AP@5  & 0.384 & 0.843  & \textbf{0.960}              \\ 
\hline
    & Spearman's    & 0.480 & 0.904  & \textbf{0.957}   \\
    & Kendall's tau & 0.372  & 0.766  & \textbf{0.858}   \\
    & AP@10 & 0.588 & 0.931 & \textbf{0.966} \\
\multirow{-4}{*}{Taste}  & AP@5  & 0.457  & 0.960              & \textbf{0.955}    \\ 
\hline                         
\end{tabular}
\end{spacing}
\end{table}

\subsection{Quality of discovered topics}
\vspace{-5pt}
We also compare the performance of different models in terms of topic findings that are measured by several quantitative metrics commonly used in topic models: 
(1) Coherence score for top 10 words of each topic. 
(2) Held-out likelihood. We perform a 5-fold cross validation on the experimental corpora and report their average results. 
(3) Hierarchical affinity for measuring the goodness of a topic hierarchy.

Table \ref{tab-case-2} summarizes the topic quality metrics from different models on the two corpora.
It is seen that MH-STM achieves the best coherence on both corpora, and it exceeds the SHLDA model in discovered hierarchical affinity. 
The held-out likelihoods of MH-STM outperform the flat model of SLDA but slightly fall behind the SHLDA on both corpora, which is consistent with the simulation observations in Table~\ref{tab-simul-2}.

\begin{table}[ht]
\begin{spacing}{1.2}
\centering
\caption{Topic quality evaluations on two corpora}
\label{tab-case-2}
\begin{tabular}{ccccc}
\hline
Dataset & Metrics   & SLDA   & SHLDA                    & MH-STM              \\ 
\hline
& Coherence $\uparrow$ & -3.140 & -3.885 & \textbf{-3.134} \\
& Held-out likelihood $\uparrow$  & -5.640 & \textbf{-5.241}  & -5.246  \\
\multirow{-3}{*}{Beer} & Hierarchical Affinity $\uparrow$ &     ---   & 2.247 & \textbf{4.804}            \\
\hline
& Coherence $\uparrow$ & -3.306 & -3.184 & \textbf{-3.068} \\
& Held-out likelihood $\uparrow$  & -6.164  & \textbf{-5.974} & -6.021  \\
\multirow{-3}{*}{Laptop} & Hierarchical Affinity $\uparrow$ &  --- &  1.450 & \textbf{4.639} \\
\hline
\end{tabular}
\end{spacing}
\end{table}

\subsection{Qualitative analysis on topic hierarchy}
\begin{figure}[!htp]
	\centering
	\setlength{\abovecaptionskip}{2pt}
	\subfigure[SHLDA]{
	    \label{fig-case-1a}
	    \begin{minipage}{0.97\linewidth}
	    \centering	    
	    \setlength{\abovecaptionskip}{2pt}
	    \includegraphics[scale=0.3]{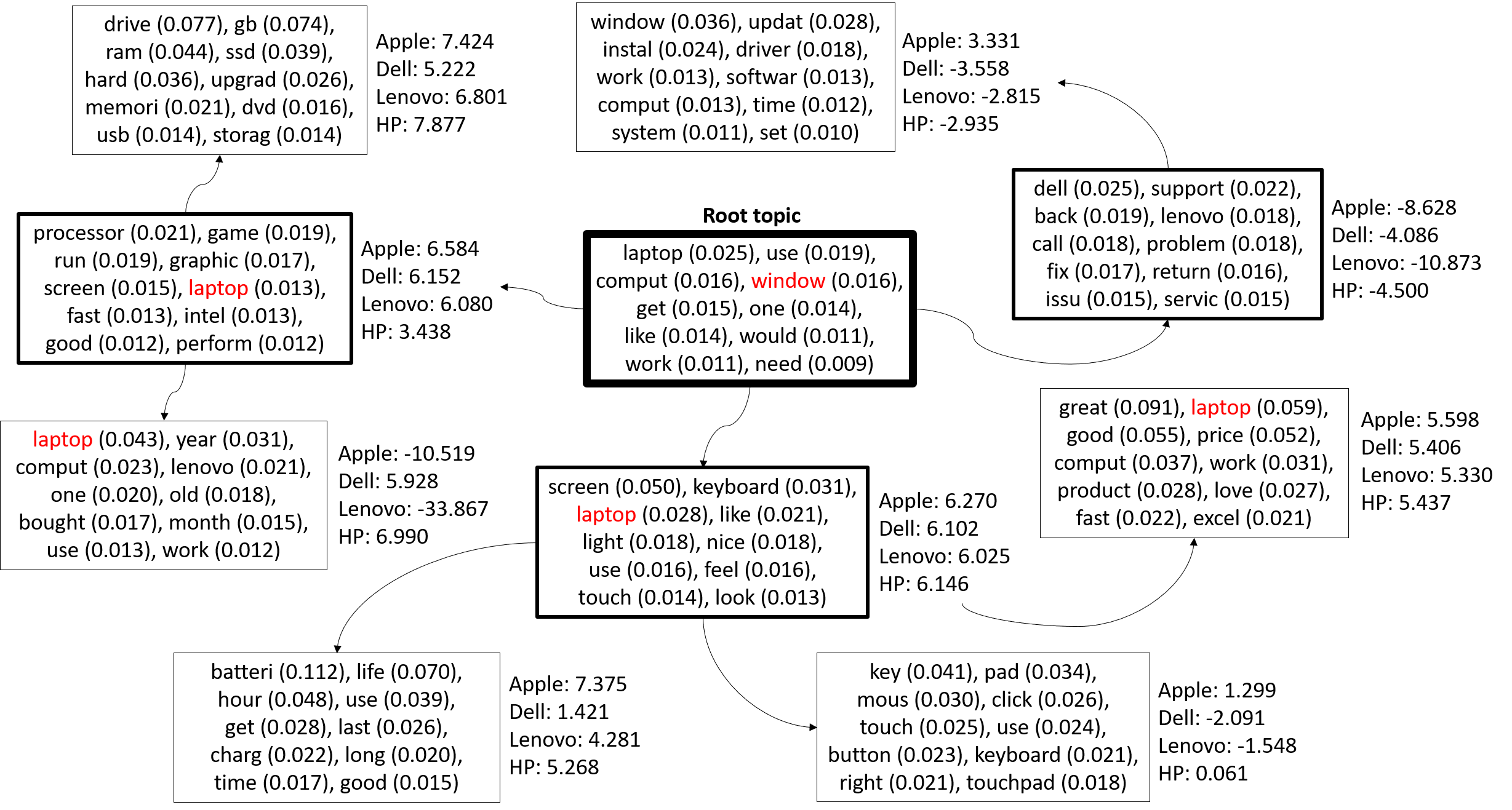}
	    \end{minipage}
	}    
	\subfigure[MH-STM]{
	    \label{fig-case-1b}
	    \begin{minipage}{0.97\linewidth}
	    \centering	    			\includegraphics[scale=0.3]{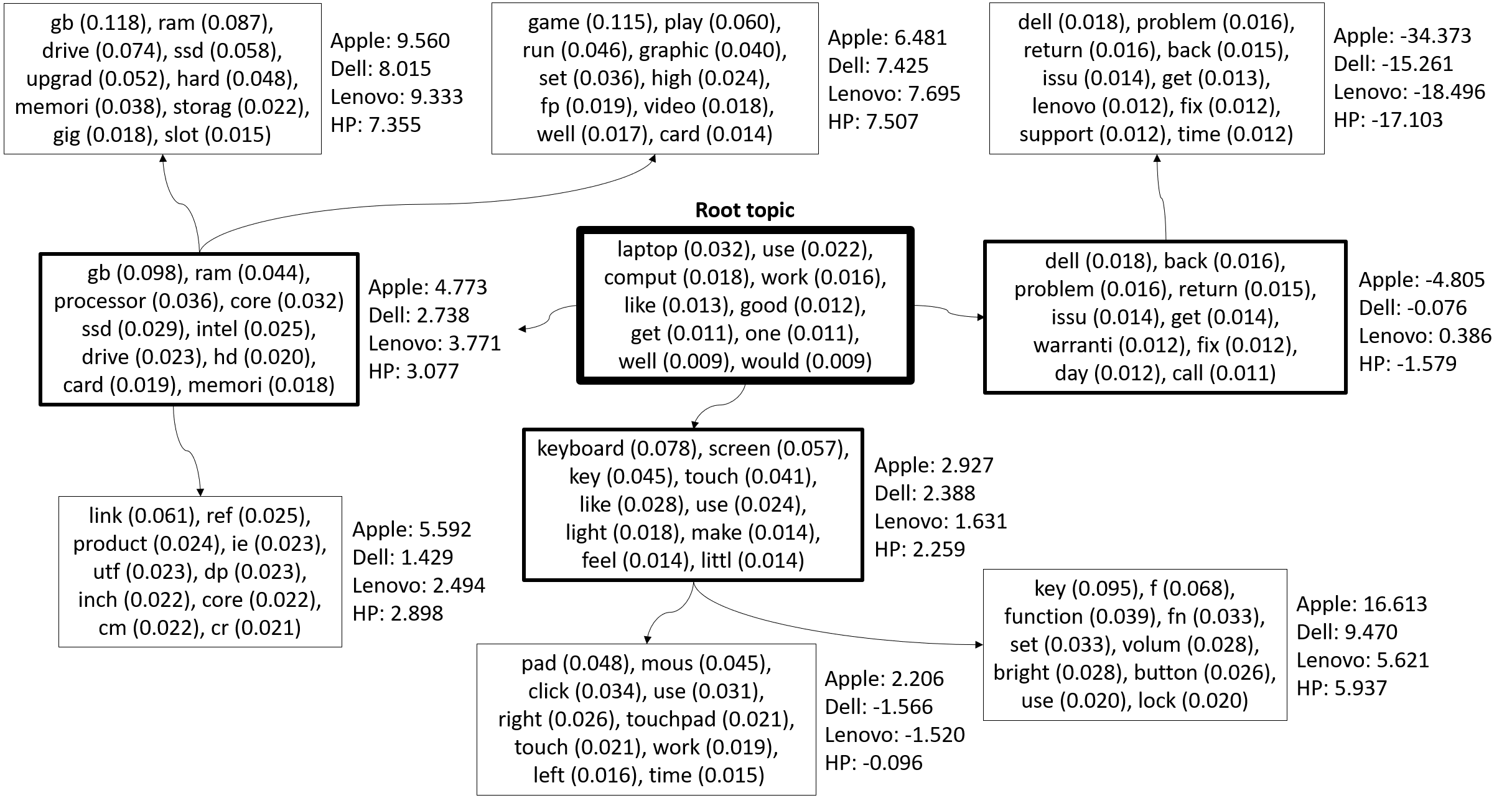}
	    \end{minipage}
	} 
	\caption{ Portions of 3-level topic hierarchy discovered from laptop reviews by SHLDA and MH-STM. Each topic is represented by its top 10 words with corresponding word probabilities under the topic. The numbers on the right side of each topic are brand-specific regression parameters learned with that topic. }
	\label{fig-case-1}
\end{figure}

It is also interesting to visualize the topic hierarchies obtained from the proposed method.
Figure \ref{fig-case-1} illustrates portions of 3-level topic hierarchies discovered from the laptop corpus respectively by SHLDA and MH-STM, where the top words with their probabilities and the brand regression parameters specific to each topic are presented.   
Both hierarchies present a general-to-specific tendency that root topics are general, and topics close to the leaves are more specific. Moreover, the regression parameters close to leaf topics learned by both models present overall higher variations among different brands, indicating that customers' polarities towards different brands are more differentiated in lower-level topics. One can obtain aspect-specific brand rankings by referring to the brand regression parameters of that aspect. 

One can see that MH-STM provides a better hierarchical topic-word association in comparison with SHLDA.
For example, the general word ``laptop'' is mixed in each level of the hierarchy by SHLDA, while it is concentrated on the root topic by MH-STM. 
On the other hand, the aspect-specific words are more assigned with lower-level topics by MH-STM. For instance, the top word ``window'' under the root topic by SHLDA is degraded in the root topic by MH-STM. Overall, this qualitative comparison result demonstrates the effectiveness of the proposed HPU scheme, in which words are assigned with different weights along the hierarchy, forcing topic-word associations to coincide with their general-to-specific characteristics.

\subsection{Runtime comparison}
\begin{figure}[ht]
\centering
\includegraphics[scale=0.45]{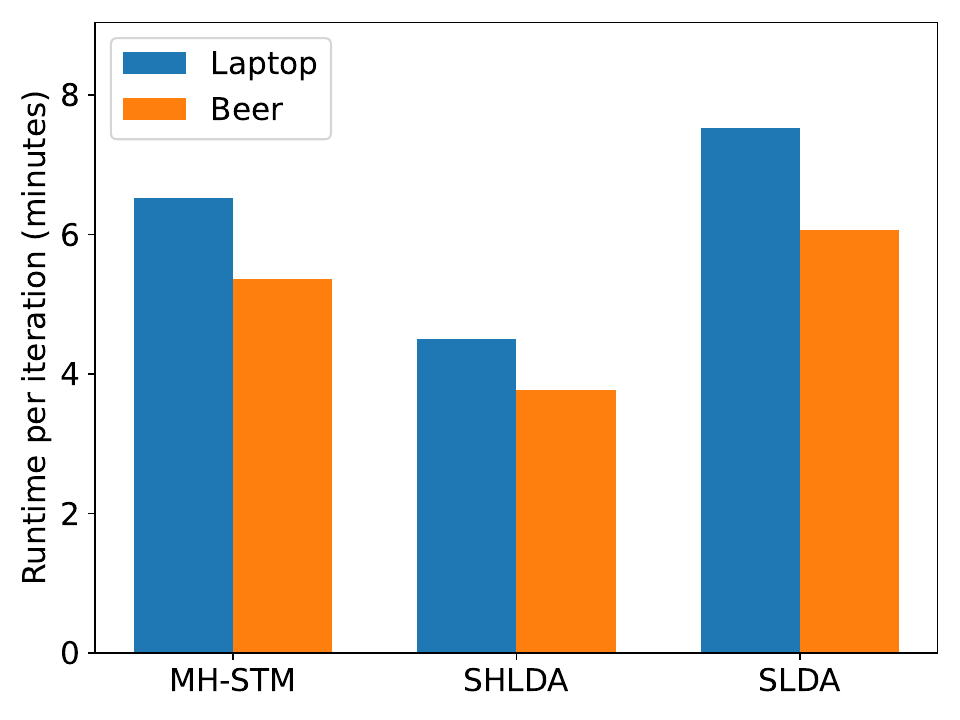}
\caption{Runtime of different methods on two datasets.}
\label{fig-case-2}
\end{figure}

The runtime of different methods on the two real-world datasets are investigated in Figure \ref{fig-case-2}.
All methods were implemented in Python 3.10 and conducted on a 64-bit Windows 11 machine with Intel Core Ultra 7 processor and 32GB memory. It shows the runtime of MH-STM ranks between the other two methods. Specifically, compared to the other hierarchical method SHLDA, the proposed model requires an additional update of weight matrix in each iteration, leading to relatively higher computational cost than the SHLDA model.

\section{Conclusion}
\label{conclusion}
In this work, proposed a multifacet hierarchical sentiment-topic model (MH-STM) for multi-aspect brand comparison automatically from online reviews. 
The proposed model is capable of extracting both hierarchically structured topics shared among various brands and topic-specific sentiment polarities across brand competitors via a unified generative framework.
Moreover, the hierarchical topic modeling is combined with a novel hierarchical P\'olya urn scheme that enhances the topic-word association by incorporating different burstiness of words among the topic hierarchy based on their general-to-specific characteristics. 

Through two real case studies from Amazon's laptop review corpus and RateBeer's beer review corpus, 
it is found that the proposed MH-STM have the merits in two major aspects. First, the extracted topics show higher coherence and more reasonable hierarchical relations.
Second, the proposed method produces more accurate brand rankings on a specified aspect.
Note that the proposed method can be easily adapted to a broad range of online review corpora in different application areas, serving as an effective tool of online brand positioning. 

There are several directions for the future work. 
It is worthwhile to consider a scalable extension of the MH-STM model. The current model applies a fully collapsed Gibbs sampler and requires an additional update of weight matrix in each iteration. 
It will be interesting to incorporate some efficient partially collapsed Gibbs sampling algorithms for hierarchical topic models \citep{chen2018scalable} to develop scalable algorithms. 
In the current simulation, we conduct model evaluation on moderate-scale topic hierarchies to achieve a balance between the efficiency and effectiveness of validation.
With a scalable extension, model evaluation can be conducted efficiently on a larger-scaled topic hierarchy with extended breadth and depth, which can better approximate the real topic hierarchies in practical applications.
Another potential extension is to handle unbalanced topic hierarchies. The current model could overlook some scarce topics when there are extremely unbalanced topic distributions. One remedy for improvement is to incorporate prior knowledge such as topic seeds \citep{jagarlamudi2012incorporating}, or uncover topics from word co-occurrence networks where the distribution over topics is less skewed \citep{zuo2016word}. 
\backmatter

\bmhead{Acknowledgements}
The authors would like to thank the editor and reviewers for their valuable comments to improve this paper.
This work was supported by the National Natural Science Foundation of China (grant number 72201212 and 72495122).

\bmhead{Data and code availability statements}
Data and code to perform the experiments in this work are publicly available at GitHub: \url{https://github.com/LiangQiao94/MH-STM}.

\setlength{\bibsep}{0.45em}

\end{document}